\documentclass[amsmath,amssymb,aps,prl,reprint,floatfix]{revtex4-1}

\usepackage{graphicx}
\usepackage{dcolumn}
\usepackage{bm}
\usepackage[none]{hyphenat}
\usepackage[mathlines]{lineno}
\usepackage{float}
\usepackage [english]{babel}
\usepackage[utf8]{inputenc}

\newcommand{\ket}[1]{\left|#1\right\rangle}

\newcommand{\broket}[3]{\left\langle #1|#2|#3\right\rangle}

\newcommand{\ketbra}[2]{\left|#1\right\rangle\left\langle#2\right|}
\newcommand{\rom}[1]{\uppercase\expandafter{\romannumeral #1\relax}}

\begin{document}

\title{Theory of robust multi-qubit non-adiabatic gates for trapped-ions}

\author{Yotam Shapira}
\email{yotam.shapira@weizmann.ac.il}
\author{Ravid Shaniv}
\author{Tom Manovitz}
\author{Nitzan Akerman}
\author{Lee Peleg}
\author{Lior Gazit}
\author{Roee Ozeri}
\affiliation{%
Department of Physics of Complex Systems, Weizmann Institute of Science, Rehovot 7610001, Israel
}%
\author{Ady Stern}
\affiliation{%
Department of Condensed Matter Physics, Weizmann Institute of Science, Rehovot 7610001, Israel
}%
\begin{abstract}
The prevalent approach to executing quantum algorithms on quantum computers is to break-down the algorithms to a concatenation of universal gates, typically single and two-qubit gates. However such a decomposition results in long gate sequences which are exponential in the qubit register size. Furthermore, gate fidelities tend to decrease when acting in larger qubit registers. Thus high-fidelity implementations in large qubit registers is still a prominent challenge. Here we propose and investigate multi-qubit entangling gates for trapped-ions. Our gates couple many qubits at once, allowing to decrease the total number of gates used while retaining a high gate fidelity. Our method employs all of the normal-modes of motion of the ion chain, which allows to operate outside of the adiabatic regime and at rates comparable to the secular ion-trapping frequency. Furthermore we extend our method for generating Hamiltonians which are suitable for quantum analog simulations, such as a nearest-neighbour spin Hamiltonian or the Su-Schrieffer-Heeger Hamiltonian.
\end{abstract}

\maketitle

\section{\rom{1}. Introduction}

Entanglement gates are at the core of universal quantum computing. The central operating paradigm of such computers is to implement quantum algorithms, i.e unitary operators acting on the qubit register, by decomposing them into a concatenation of elements of a universal gate set \cite{DiVincenzo1995,Barenco1995,Kitaev1997}. The universal gate set usually consists of arbitrary single qubit operations and a two-qubit entanglement gate, e.g a Controlled-NOT gate, which can be performed on any two qubits in the qubit register.

Trapped ion qubits are a leading platform for the realization of a universal quantum computer, already demonstrating many of the required components with outstanding fidelities \cite{Myerson2008,Harty2014,Ballance2016,Bermudez2017,Linke2017,Bruzewicz2019,Wright2019}. Entanglement gates, which are considered the bottleneck of such realizations, have recently been at the focus of many theoretical and experimental investigations aimed at improving their fidelity, efficiency and robustness \cite{Roos2008,Haddadfarshi2016,Palmero2017,Manovitz2017,Wong2017,Schafer2018,Leung2018,Webb2018,Shapira2018,Figgatt2018,Milne2018,Leung2018b,Grzesiak2019,Sutherland2019,Blumel2019,Lu2019,Sutherland2019b}.

However a multi-qubit fault-tolerant quantum computer has not been achieved yet with trapped ions, or with any other quantum platform. A central challenge hindering the appearance of such quantum computers is that of scaling-up. In particular, when the number of the quantum bits in the register increases the number of concatenated universal gate set elements increases exponentially \cite{Kitaev1997} while the fidelity of each separate element generically drops \cite{Monroe2013}. 

A possible resolution of this challenge is by expanding the universal gate set, making it over-complete, by adding different types of entanglement gates, specifically, all-to-all multi-qubit entanglement gates. It has already been shown that these multi-qubit gates can increase the fidelity of many quantum algorithms \cite{Martinez2016,Maslov2018}. 

The same methods used for creating computing-oriented entangling gates in trapped-ion systems are also used for analog spin-Hamiltonian simulations. In these simulations spin-spin interactions are generated with an interaction strength that scales as $r^{-\alpha}$, where $r$ is the distance between ions and $0\leq\alpha\leq3$ \cite{Poras2004,Islam2013,Jurcevic2017,Zhang2017}.

Here we propose and investigate a family of multi-qubit entangling gates for trapped ions. Conventionally, trapped ions entangling gates operate by coupling to a single normal-mode of motion of the ion-chain while the presence of other normal-modes limits the gate rate. Our gates purposefully couple to all normal-modes of motion of the ion-chain and can therefore operate in the non-adiabatic regime. Furthermore, the different normal-modes of motion can be used to generate a wide variety of interactions. We present examples of all-to-all entangling gates, which are especially suited for quantum computing and examples of spin-Hamiltonians such as the nearest-neighbour Hamiltonian. 

\section{\rom{2}. Main results}

Our main result is a family of multi-qubit entangling gates for trapped ion qubits, which generate a quantum evolution operator of the form $\exp\left(i\sum_{i,k=1}^N j_{i,k}\hat{\sigma}_{y,i}\hat{\sigma}_{y,k}\right)$, with $\hat{\sigma}_{y,i}$ the Pauli-$\hat{y}$ operator acting on the $i$'th qubit in the $N$ qubit register, and $j_{i,k}$ is a symmetric coupling matrix. 

Specifically we focus on equal all-to-all entanglement gates, for which $j_{i,k}^{\text{all-to-all}}=\frac{\pi}{4}$ for all $i$ and $k$, and spin-Hamiltonian couplings such as nearest-neighbour interactions, for which $j_{i,k}^{\text{n.n}}=\phi\left(\delta_{i,k+1}+\delta_{i,k-1}\right)$, with an arbitrary $\phi$. Our method, however, can be used to implement many other spin-coupling Hamiltonians.

Our method requires only global uniform interaction of a multi-tone light-field with the ions. The field spectrum is comprised of harmonics of the gate time, with a bandwidth that overlaps the frequencies of the normal-modes of motion of the ion-chain. Implementing a specific interaction type is done by choosing the relative amplitudes of the different tones. We do not require individually addressing any of the ions, and thus our method is relatively simple to implement and natural to most trapped-ion quantum processor architectures. 

This operational principle is made possible by exploiting a counter-intuitive fact about the orthogonal normal-modes of motion of the ion-crystal: the coupling matrix mediated by a linear combination of some of the normal-modes can be made to appear as if it was generated by other, orthogonal, normal-modes. Thus, instead of decoupling the different modes of motion we utilize them and generate an accumulated effect. This allows us to generate non-adiabatic entangling gates with rates comparable to the secular ion-trapping frequencies.

\begin{figure}
\includegraphics[width=\columnwidth]{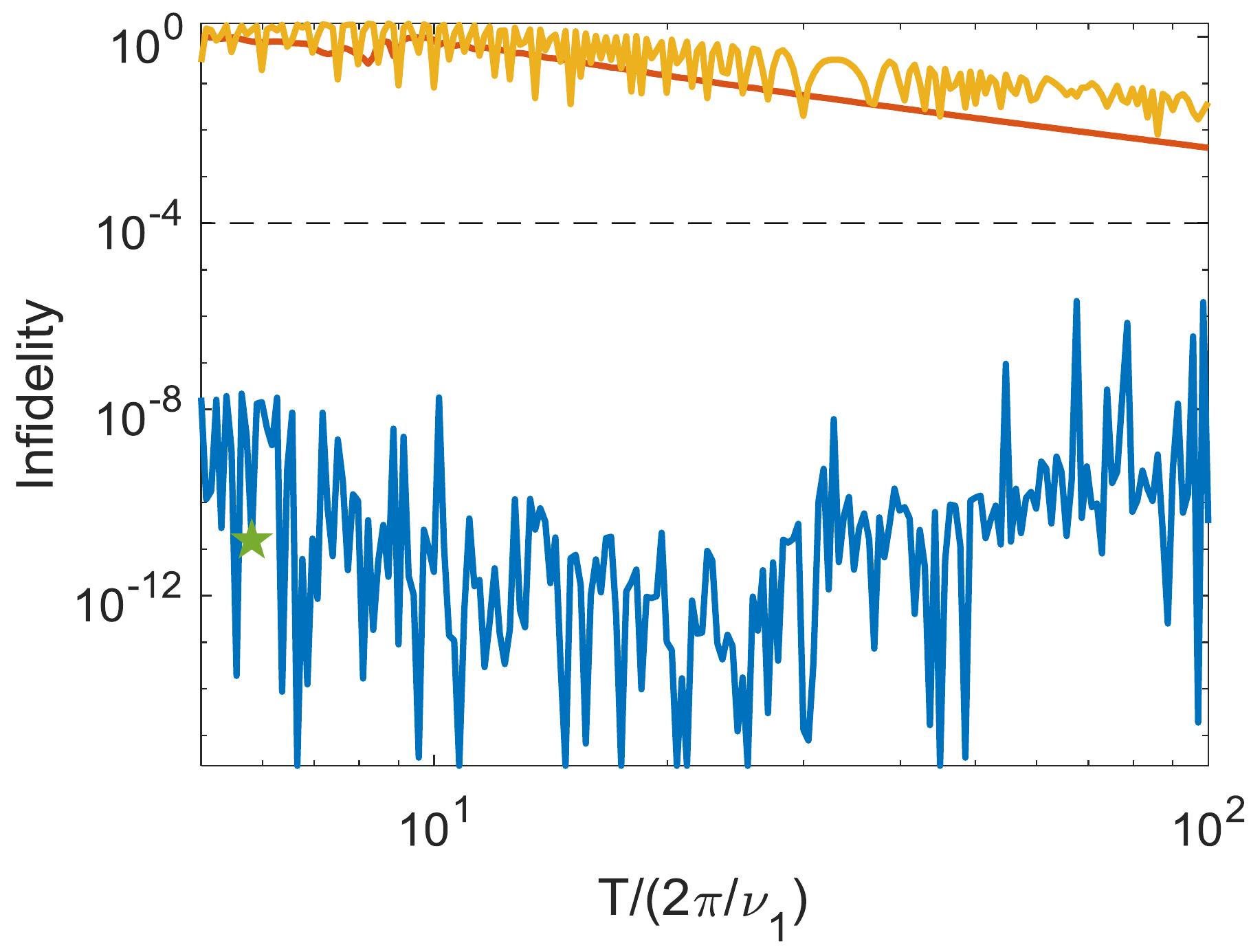}
\caption{Comparison of 6-ion multi-mode entangling gate fidelity to MS and CarNu(2,3,7) gates for varying gate times. The gate time is given in dimensionless units, with respect to the center-of-mass axial mode period $\frac{2\pi}{\nu_1}$. An example gate, with gate time $T\approx5.8\frac{2\pi}{\nu_1}$ is highlighted (green star) and analyzed with more detail below. We have designed our gate such that the infidelity is lower than $10^{-4}$ (dashed black). Indeed our gate (blue) performs well. However the MS (yellow) and CarNu(2,3,7) (red) gates, acting on the axial center-of-mass mode, fail to generate a high-fidelity operation as they are operating outside of their adiabatic regime.}\label{figFidel6}\end{figure}

As we show below, our all-to-all gates do not require the full knowledge of the amplitudes of each of the $i=1,...,N$ ions in each of the $j=1,...,N$ normal-modes. We only need to know the normal-mode frequencies. Furthermore the laser power overhead required to implement our gates is small.

The expected infidelity of all-to-all entanglement gates scales as $1-F\sim\frac{T}{T_2}N^\alpha$, with the gate time $T$, the single-qubit decoherence time $T_2$ and $1\leq\alpha\leq2$ \cite{Bermudez2017}, depending on realization, error-model and initial state \cite{Monz2011,Ozaeta2019}. Thus operating at high-rates is crucial for scaling-up the qubit register.

In addition we endow our gates with robustness properties that makes them resilient to many types of errors, such as pulse-timing errors, trap secular frequency drifts, optical phase drifts (relevant to Raman configurations), normal-mode heating among other examples.

\begin{figure}[t]
\includegraphics[width=\columnwidth]{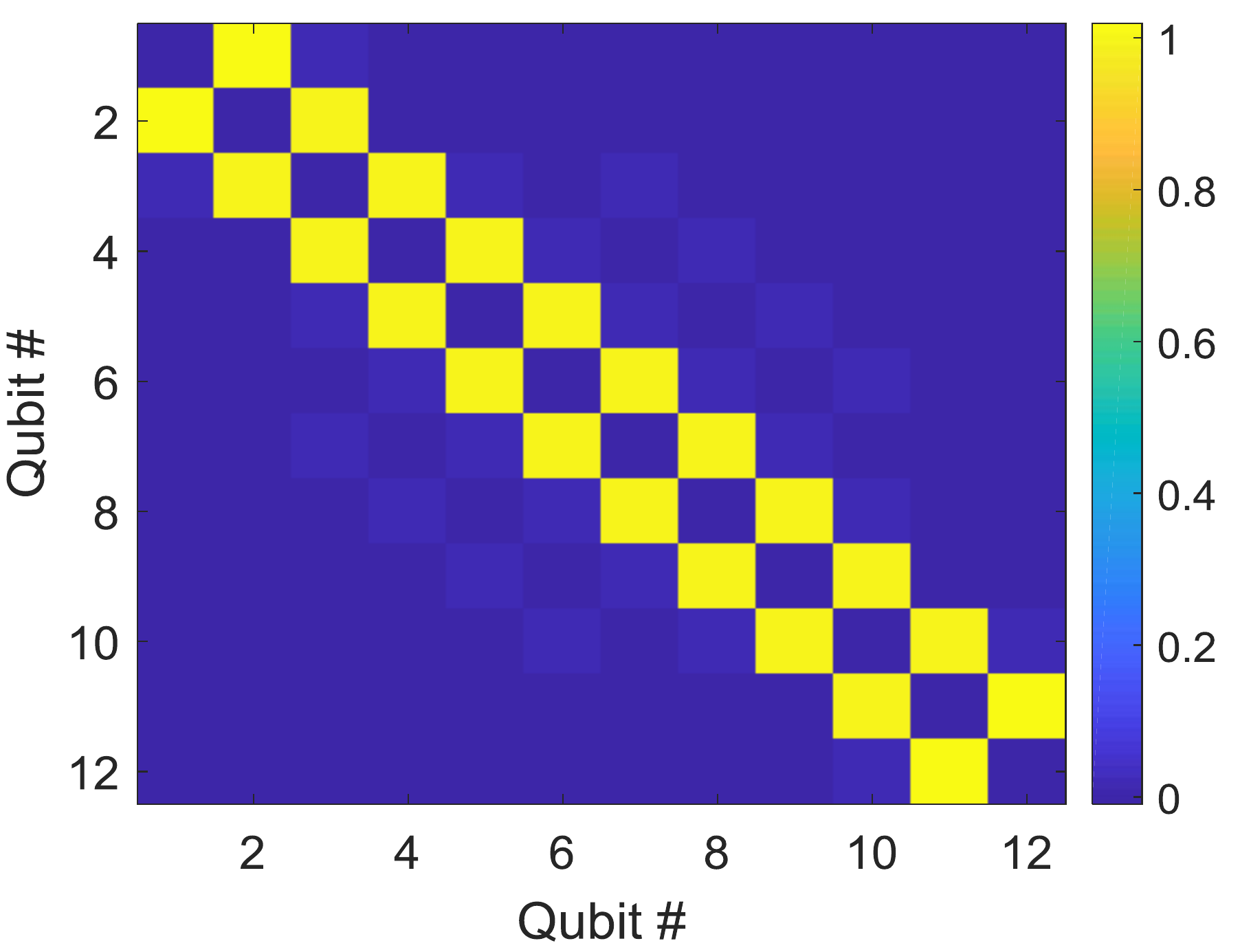}
\caption{Example of the coupling matrix of an entangling gate realizing a nearest-neighbour interaction Hamiltonian. The gate is designed such that the resulting coupling matrix is $j_{i,k}=\phi\left(\delta_{i,k+1}+\delta_{i,k-1}\right)$, where $\phi$ is a coupling strength. Here the realization fidelity is better than 0.999.}\label{figMatNN}\end{figure}

Before diving into the details of our method, we show examples for the couplings and the entanglement fidelity that can be achieved with our scheme in two figures.

Figure \ref{figFidel6} shows simulation results for different all-to-all entanglement gates, acting on a $N=6$ qubit register in a harmonic ion-trap, for varying gate rates. We benchmark our gate by its fidelity of rotating the qubit ground state to a Greenberger–Horne–Zeilinger (GHZ) \cite{Greenberger1989} state, since GHZ states are good indicators to coherent gate errors \cite{Gottesman2019}. We compare our gate's performance to previously demonstrated methods, such as the M\o lmer–S\o rensen gate \cite{Molmer11999,Sorensen2000} (MS) and the CarNu(2,3,7) gate \cite{Shapira2018} that are using a single mode of motion. The multi-ion multi-mode gates (blue) exhibits low infidelity, which is clearly separated from the MS (yellow) and CarNu (red) gates, operating at a much higher infidelity due to their coupling to unwanted motional modes and to the carrier transition.

Figure \ref{figMatNN} exemplifies how our method is used for generating spin-Hamiltonians for analog quantum simulations.  It shows a simulation of the nearest-neighbour coupling matrix $j_{i,k}$ we implemented, on a $N=12$ qubit register. The nearest-neighbour structure is clearly seen. Indeed the overlap between the simulated $j_{i,k}$ and $j_{i,k}^{\text{n.n}}$ is better than 0.999. Below we show further examples of other spin models such as next-nearest neighbour and the Su-Schrieffer-Heeger model \cite{Su1980}.

\section{\rom{3}. All-to-all entanglement gate derivation}
We begin by deriving the system Hamiltonian. The non-interacting lab-frame Hamiltonian of $N$ trapped ions is,
\begin{equation}
    \hat{H}_0=\sum_{k=1}^{N}\left(\hbar\nu_{k}\left(\hat{a}_{k}^{\dag}\hat{a}_{k}+\frac{1}{2}\right)+\frac{\hbar\omega_{0}}{2}\hat{\sigma}_{z,k}\right). \label{eqH0}
\end{equation}
such that $\hat{a}_j$ is the lowering operator of the $j$'th normal-mode of motion with frequency $\nu_j$, $\omega_0$ is the single qubit separation frequency and $\hat{\sigma}_{z,k}$ is the Pauli-$\hat{z}$ spin-operator acting on the $k$'th qubit. 

Here we make use of the normal-modes of motion along a single direction, and implicitly assume that modes of the other directions are decoupled from the evolution. However our derivations below are easily generalized to the complete set of $3N$ normal-modes.

The ions are driven by a multi-chromatic laser field, containing $2M$ frequencies arranged in pairs, $\left\{\omega_0\pm\omega_i\right\}_{i=1}^M$. Each component has phase $\phi_{\pm,i}=\pm\phi_i$, i.e the average phase of each pair is $0$, and each pair has the same amplitude $\Omega r_i$, with $\Omega$ a characteristic Rabi frequency and $r_i\in\mathbb{R}$ (such that $r_i\rightarrow -r_i$ is the same as $\phi_i\rightarrow\phi_i+\pi$). In total this driving field is determined by the $3M$ degrees of freedom, $\boldsymbol{\omega}$, $\boldsymbol{\phi}$ and $\boldsymbol{r}$. 
The resulting interaction due to this field is,
\begin{equation}
    \hat{V}=2\hbar\Omega\sum_{i=1}^{M}r_{i}\sum_{n=1}^{N}\hat{\sigma}_{x,n}\cos\left(k\hat{x}_{n}-\omega_{0}t\right)\cos\left(\omega_{i}t+\phi_{i}\right),\label{eqV}
\end{equation}
where $\hat{\sigma}_{x,n}$ is a Pauli-$\hat{x}$ spin-operator acting on the $n$'th qubit, $k$ is the laser momentum vector projected on the normal-mode direction of motion and $\hat{x}_n$ is the position operator of the $n$'th qubit. The wave vectors $k$ are approximately identical for all frequencies. We note that we assumed implicitly that the ions are driven with a uniform global field, i.e $\Omega$ has no $n$-index. 

Changing to an interaction picture with respect to $\hat{H}_0$, performing an optical-frequency rotating wave approximation and performing the Lamb-Dicke approximation (see appendix \rom{1}), we obtain,
\begin{equation}
    V_{I}=\hbar\Omega\sum_{j=1}^{N}\left(f_{j}\left(t\right)\hat{q}_{j}+g_{j}\left(t\right)\hat{p}_{j}\right)\hat{J}_{y,j}, \label{eqVI5}
\end{equation}
with $f_{j}\left(t\right)+ig_j\left(t\right)=\frac{2\sqrt{2}}{\sqrt{N}}\eta_{j}\sum_{i=1}^{M}r_{i}\cos\left(\omega_{i}t+\phi_{i}\right)e^{i\nu_{j}t}$, $\hat{q}_j$ ($\hat{p}_j$) is the dimensionless position (momentum) operator associated with the $j$'th normal-mode of motion, $\eta_j\equiv k\sqrt{\frac{\hbar}{4\pi m\nu_j}}$ is the Lamb-Dicke parameter of the $j$'th normal-mode. The spin coupling operator is $\hat{J}_{y,j}=\frac{\sqrt{N}}{2}\sum_{n=1}^{N}O_{j,n}\hat{\sigma}_{y,n}$, such that $O_{j,n}$ is the normalized participation of the $n$'th ion in the $j$'th mode of motion. It is a generalization of the global rotation operator, $\hat{J}_y=\frac{1}{2}\sum_{n=1}^N \hat{\sigma}_{y,n}$. Eq. \eqref{eqVI5} is lacking a carrier-coupling term, which has been omitted. We justify this omission below.

For harmonic confinement (along the axial or the radial directions) we designate the center-of-mass mode as mode number 1, and denote $\hat{J}_{y,1}=\hat{J}_y$. In-order to implement all-to-all entanglement gates we require no explicit knowledge of $O$.

Equation \eqref{eqVI5} is the non-adiabatic, multi-ion, multi-mode, multi-tone generalization of Eq. (6) of Ref. \cite{Sorensen2000}. As such it follows an analogous solution, that is,
\begin{equation}
\begin{split}
    \hat{U}&=\prod_{j=1}^{N}\left(e^{-iA_{j}\hat{J}_{y,j}^{2}}e^{-iF_{j}\left(t\right)\hat{q}_{j}\hat{J}_{y,j}}e^{-iG_{j}\left(t\right)\hat{p}_{j}\hat{J}_{y,j}}\right) \\
    \alpha_j\left(t\right)&\equiv F_j\left(t\right)+iG_j\left(t\right)=\int\limits _{0}^{t}dt^{\prime}\left(f_{j}\left(t^\prime\right)+ig_j\left(t^\prime\right)\right) \\
    A_{j}\left(t\right)&=\int dt^{\prime}F_{j}\left(t^{\prime}\right)\frac{dG_{j}\left(t^{\prime}\right)}{dt^{\prime}}.
\end{split}\label{eqU}
\end{equation}
The evolution operator in Eq. \eqref{eqU} shows that the system evolution in the $j$'th normal-mode phase space is along the curve $\alpha_j\left(t\right)$. The operator product in Eq. \eqref{eqU} is well defined since the operators associated with different normal-modes commute, thus no ordering is required.

Assuming that at the gate time all trajectories return to $0$, i.e $\alpha_j\left(t=T\right)=0$, then at this time, the evolution operator can be written as exclusively acting in the qubit subspace and is determined by a sum of mode-dependent entangling operators, $\hat{J}_{y,j}^2$, with a phase proportional to the area, $A_j\left(T\right)$, enclosed by the phase-space trajectory of mode $j$. We define $\varphi_j=A_j\left(T\right)$ as the mode-dependent entangling phase. A natural scaling of the necessary drive power with the number of ions can be predicted by noticing that the $A_j$'s are proportional to $\Omega^2/N$. We therefore expect $\Omega\propto\sqrt{N}$.

We next derive general constraints on the entangling phases $\left\{\varphi_j\right\}_{j=1}^N$ such that a desired multi-qubit entangling gate is formed. 
For an all-to-all coupling gate, an obvious method to rotate the ground state to a GHZ state is by demanding that $\alpha_{j=1,...,N}\left(T\right)=0$, $\varphi_{j\geq2}=0$ and $\varphi_1=\frac{\pi}{2}$. That is, the entangling operation can be obtained by enclosing an area of $\frac{\pi}{2}$ in the center-of-mass phase-space while not accumulating any area in all other modes of motion. This is precisely what is achieved in Ref. \cite{Molmer21999} in the adiabatic regime. 

We would like to obtain the same end result, but in the non-adiabatic regime. Thus we ask whether the condition $\varphi_{j\geq2}=0$ is necessary. Surprisingly the answer is no, and it may be replaced by a significantly less restrictive constraint. Specifically we use the relation,
\begin{equation}
    \textbf{1}=e^{i\sum_{j=1}^N \hat{J}_{y,j}^2}\Rightarrow e^{i\hat{J}_{y,1}^2}=e^{-i\sum_{j=2}^N \hat{J}_{y,j}^2},\label{eqJSum}
\end{equation}
which shows that when all of the $j\geq2$ modes are equally coupled, then a center-of-mass-like effect is generated, with opposite coupling. Thus the necessary condition is in fact, $\varphi_1-\varphi_{j\geq2}=\frac{\pi}{2}$ for all $j\geq2$. This does not merely reduce the number of constraints on $\varphi_j$, but also allows for non-vanishing entanglement phases associated with all normal-modes of motion.

Equation \eqref{eqJSum} above is non-intuitive, as it shows that a sum over the spin-couplings of orthogonal modes can generate that of a different orthogonal mode. This is of course only valid since the summation is over the operators squared, $\hat{J}_{y,j}^2$ (mode orthogonality would prohibit a similar identity for the $\hat{J}_{y,j}$'s). We prove this identity in appendix \rom{2}.

The only knowledge of the normal-modes structure we used is that the first mode is a center-of-mass mode. As we show below, this means that in order to generate an all-to-all entangling gate we only need to know the frequencies of the remaining modes, as they determine the different Lamb-Dicke parameters, but not the specific participation of the $i$'th ion in the $j$'th normal-mode, $O_{j,i}$. 

The identity in Eq. \eqref{eqJSum} can be used not only for all-to-all type couplings, but also to efficiently generate other types of couplings such as the nearest-neighbour interaction shown in Fig. \ref{figMatNN}, and for general interactions which can be written as linear combination of the $\hat{J}_{y,j}^2$ operators, even when a center-of-mass mode doesn't exist.

The driving field acts between time $t=0$ and the gate time $t=T$. Furthermore, we show below that it is beneficial to use a drive that vanishes continuously at its edges. Such drives can always be expanded in a Fourier-sine basis in harmonics of $\frac{2\pi}{T}$. Thus we fix $2M$ degrees of freedom of the driving field such that $\omega_n=\frac{2\pi}{T}n$ and $\phi_n=\frac{\pi}{2}$ for $n=1,...,M$. Choosing a harmonic basis for the gate drive has already been proven useful in several entangling gate schemes \cite{Palmero2017,Shapira2018,Blumel2019}.

This approach eliminates the need to optimize $\boldsymbol{\omega}$ and $\boldsymbol{\phi}$, and hinges all of the gate properties on the optimization of $\boldsymbol{r}$. However it comes at a price - the basis is infinite. Practically we truncate the series of tones such that all spectral components are in the vicinity of the motional modes. This is reasonable since tones that are far away from all of the normal-mode frequencies couple almost uniformly to all modes, and therefore, due to Eq. \eqref{eqJSum}, cannot significantly contribute to the gate's performance. 

This basis also highlights the speed-limit of our method. For harmonic confinement in the $N\gg1$ case, the axial-modes frequency difference between adjacent modes approaches $\frac{\nu_1}{2}$. Due to the identity in Eq. \eqref{eqJSum}, it is beneficial to place the driving frequencies between the different motional modes. However for $T<\frac{\pi}{\nu_1}$ it is no longer possible to do so, leading to a diverging drive power. 

As stated above, in order to implement our gates we must satisfy the constraint $\alpha_j\left(T\right)=0$ for all $j=1,...,N$. That is, at the gate time all phase-space trajectories return to their initial coordinates such that a state which initially had spin and motion degrees of freedom disentangled, remains disentangled after the gate operation.

Using Eq. \eqref{eqU} we note that this constraint is linear in $\boldsymbol{r}$ and can be separated to a real and imaginary part, thus it can be written as a linear relation
\begin{equation}
    L\boldsymbol{r}=\boldsymbol{0},\label{eqLinCons}    
\end{equation}
with $L=L\left(\boldsymbol{\omega},\boldsymbol{\phi}\right)$ a $2N\times M$ matrix, whose elements are,
\begin{equation}
\begin{cases}
    L_{j,i}\propto\int\limits _{0}^{T}dt \cos\left(\omega_{i}t+\phi_{i}\right)\cos\left(\nu_{j}t\right) & 1\leq j\leq N\\
    L_{j,i}\propto\int\limits _{0}^{T}dt \cos\left(\omega_{i}t+\phi_{i}\right)\sin\left(\nu_{j}t\right) & N+1\leq j\leq 2N
\end{cases},\label{eqTrajCons}
\end{equation}
with $i=1,...,M$. 

\begin{figure}
\includegraphics[width=\columnwidth]{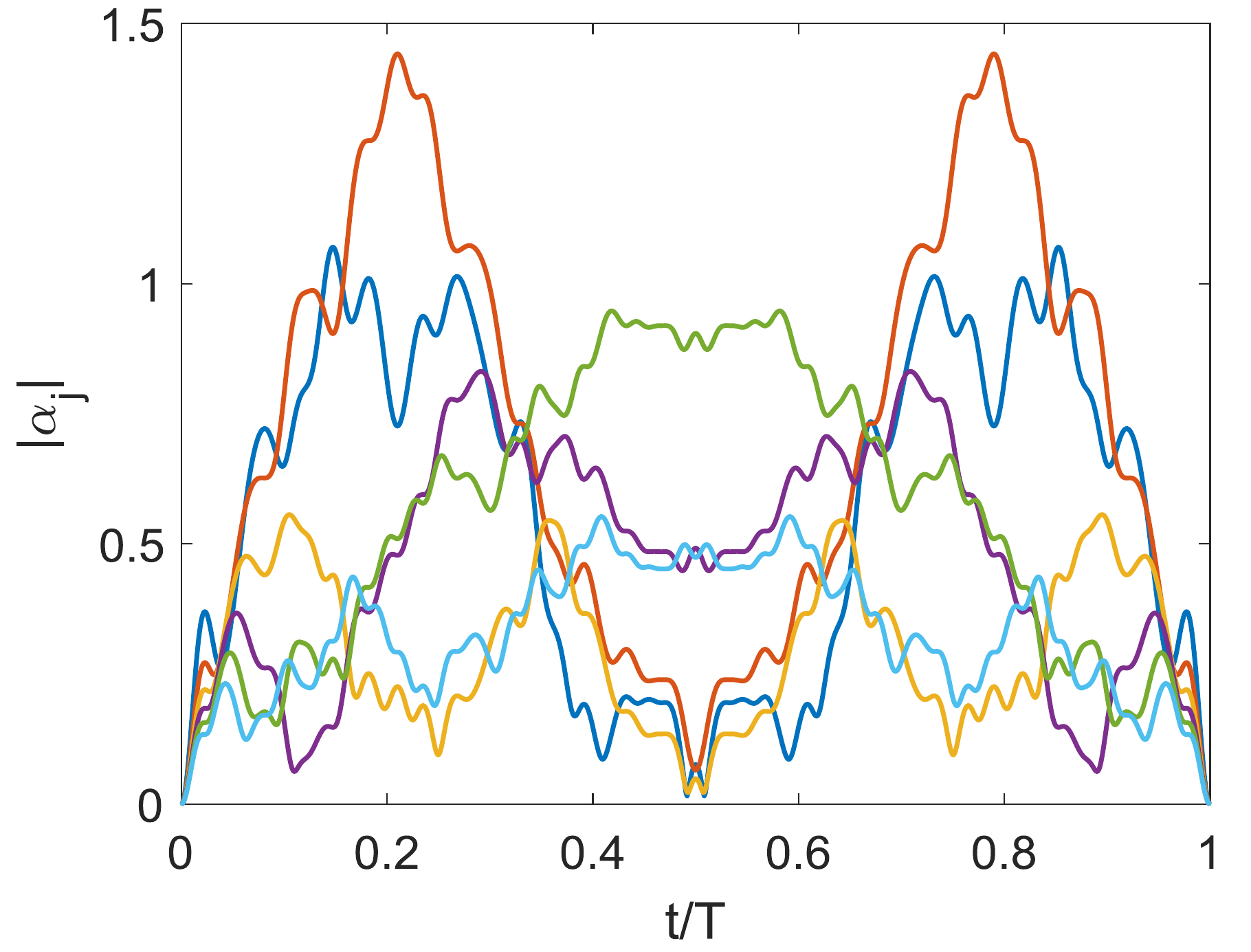}
\caption{Phase space trajectories distance of different motional modes from the origin, $\left|\alpha_j\left(t\right)\right|=\sqrt{F_j^2+G_j^2}$, during the gate operation of the highlighted example gate in Fig. \ref{figFidel6}. The figure shows the first (dark-blue), second (red), third (yellow), fourth (purple), fifth (green) and sixth (light-blue) modes. All trajectories start and end at the origin indicating that the motion is disentangled from spin degrees of freedom at the gate time.}\label{figPhaseSpace}\end{figure}

We demonstrate the different aspect of the derivation using the $N=6$ ions gate highlighted in Fig. \ref{figFidel6} (green star) as an example. The methods used to calculate the gate are provided below. Figure \ref{figPhaseSpace} shows the magnitude of the phase-space trajectories, $\left|\alpha_j\left(t\right)\right|$, of the highlighted gate, as the system evolves. Clearly all six trajectories start at $0$ at $t=0$ and end at $0$ at $t=T$ as well, indicating that the linear constraints are met. 

In addition to this linear relation, we may require that the entangling gate operation will be robust against various types of experimental imperfections and noise. Examples include pulse timing errors, normal-mode frequency drifts, normal-mode heating, optical phase noise (relevant to Raman configurations) and  non-smooth effects. Such robust gates have been previously analyzed in a similar context \cite{Roos2008,Shapira2018,Webb2018,Leung2018,Schafer2018}, and are all linear in $\boldsymbol{r}$ in any order of correction. Thus they can be incorporated as additional rows of $L$. The exact form of each of these properties is provided in appendix \rom{3}.

A particular imperfection that can be overcome by adding linear constraints is that of off-resonance carrier coupling, justifying the omission of the carrier-coupling term in deriving Eq. \eqref{eqVI4}. To do so we rewrite the Hamiltonian in Eq. \eqref{eqVI3} as the sum of the non-commuting terms, $\hat{V}_I=\hat{H}_{c.c}+\hat{H}_{MS}$, with $\hat{H}_{c.c}=\hbar\Omega\sum_{i=1}^{M}r_{i}\cos\left(\omega_{i}t+\phi_{i}\right)\hat{J}_{x,1}$ and with $\hat{H}_{MS}$ given by Eq. \eqref{eqVI5}. We make use of a Magnus expansion in order to derive constraints for the elimination of contributions of the unwanted $\hat{H}_{c.c}$ term to the evolution  \cite{Magnus1954,Roos2008,Blanes2009} (see appendix \rom{4}). This yields an additional linear constraint, $\sum_{i=1}^{M}r_{i}\cos\left(\omega_{i}T+\phi_{i}\right)=0$, which can be added to the rows of $L$. The next order contribution due to the carrier-coupling terms are quadratic in $\boldsymbol{r}$ and are treated below.

We define $K\equiv\text{null}\left(L\right)$, as a $M\times l$ matrix, the columns of which, $\boldsymbol{r}_i$, form an orthogonal basis of the null space of $L$, i.e satisfy $L\boldsymbol{r}_i=\boldsymbol{0}$ for $i=1,...,l$. Every linear combination, $\boldsymbol{r}=\sum_i\boldsymbol{r_i}$ satisfies all the linear constraints above. The linear constraints can be met only if we have a sufficient number of tones, i.e $M$ has to be larger than the number of rows of $L$.

The linear constraints guarantee that the trajectories are closed, but do not fix the entangling phases implemented by the trajectory. The entangling phases, $\varphi_j=A_j\left(T\right)$, are quadratic in $\boldsymbol{r}$. Thus, in a similar fashion to the linear constraints above, they can be written as a bi-linear form, $\varphi_j=\boldsymbol{r}^T\tilde{A_j}\boldsymbol{r}$, with the $N$ symmetric $M\times M$ matrices, whose elements are,
\begin{equation}
\begin{split}
    \left(\tilde{A}_j\right)_{i,k}= & -4\eta_{j}^{2}\int\limits _{0}^{T}dt\int\limits _{0}^{t}dt^{\prime}\bigg[\sin\left(\nu_{j}t\right)\cos\left(\nu_{j}t^{\prime}\right)\\
    & \cdot\left(\cos\left(\omega_{k}t+\phi_{k}\right)\cos\left(\omega_{i}t^{\prime}+\phi_{i}\right)+i\leftrightarrow k\right)\bigg].
\end{split}\label{eqAj}
\end{equation}

In order to restrict $\boldsymbol{r}$ to satisfy the linear constraints above and such that the entangling phase constraints in Eq. \eqref{eqJSum} are satisfied as well, we define,
\begin{equation}
    \tilde{C}_j\equiv K^T\left(\tilde{A}_1-\tilde{A}_j\right)K,\text{  }j=2,...,K,\label{eqCj}.
\end{equation}
Here, each of the $N-1$ different $\tilde{C}_j$'s is a $l\times l$ matrix. 

Thus, to find a solution to the desired phases within the null-space of $L$, the problem is reduced to choosing an $l$-element real vector, $\boldsymbol{x}$, such that the constraint,
\begin{equation}
    \boldsymbol{x}^T\tilde{C}_j\boldsymbol{x}=\varphi_1^\text{desired}-\varphi_j^\text{desired},\text{  }\forall j=2,...,N\label{eqAmpCons}, 
\end{equation}
is satisfied, where $\varphi_j^\text{desired}$ are the entanglement phases which implement the desired interaction. For an all-to-all entangling gate the r.h.s of Eq. \eqref{eqAmpCons} is given by $\varphi_1^\text{desired}-\varphi_j^\text{desired}=\frac{\pi}{2}$.

\begin{figure}
\includegraphics[width=\columnwidth]{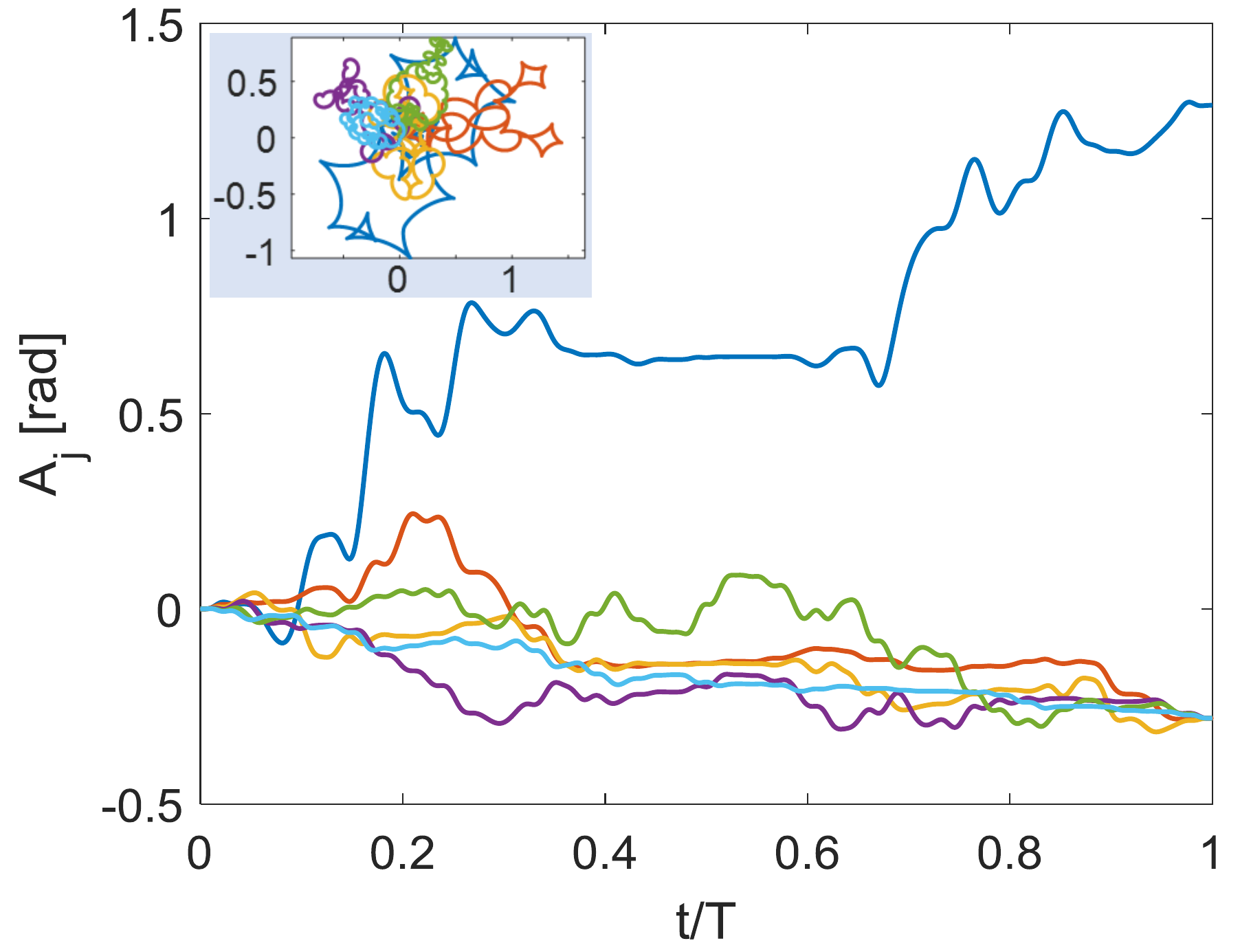}
\caption{Entangling phases of highlighted example gate in Fig. \ref{figFidel6}. Each entangling phase evolves independently from the other, however at gate time the difference between $\varphi_1$ (dark-blue) and the remaining phases, $\varphi_2$ (red), $\varphi_3$ (yellow), $\varphi_4$ (purple) $\varphi_5$ (green) and $\varphi_6$ (light-blue) is exactly $\frac{\pi}{2}$. Together with the closure of phase space trajectories, shown in Fig. \ref{figPhaseSpace}, the unitary fidelity of this gate is 1. Robustness to timing-errors is evident as all the entanglement phase curves flatten near the gate time. The inset shows the entire phase-space trajectories formed, clearly all trajectories start and end at the origin, as is evident in Fig. \ref{figPhaseSpace} as well.}\label{figPhases}\end{figure}

Figure \ref{figPhases} shows the entangling phases evolution for the $N=6$ ions gate highlighted in Fig. \ref{figFidel6}. Clearly each phase evolves seemingly independently, however at gate time the distance between the center-of-mass mode phase (blue) and the remaining is $\frac{\pi}{2}$, indicating a valid solution of Eq. \eqref{eqAmpCons} above.

For arbitrary matrices $\tilde{C}_j$'s in Eq. \eqref{eqAmpCons}, finding solutions for the naively looking Eq. \eqref{eqAmpCons} above is in fact a NP-hard problem, known as the multivariate quadratic problem \cite{Garey1979,Grenet2010}. However the "hardness" is in terms of the matrix dimension, $l$. Thus it is critical to choose $M$ such that the resulting null space dimension, $l$, is compatible with the number of quadratic constraints, i.e such that $l=\mathcal{O}\left(N\right)$. As we show below, provided an appropriate initial guess, a local numerical search yields, in most cases, satisfactory solutions, and thus the hardness of the problem does not hinder finding suitable gates for a moderate number of 10's of ions.

For the case $N=2$ ions the problem is easily solveable. A solution is formed by choosing arbitrary amplitudes, $\boldsymbol{r}$, that satisfy the linear constraints (which is numerically easy). Since there is only a single quadratic condition, $\tilde{C}_2$, then by choosing a normalization for $\boldsymbol{r}$ such that $\boldsymbol{x}^T\tilde{C}_2\boldsymbol{x}=\frac{\pi}{2}$ all constraints are met. Thus generating fast two-qubit entangling gates is conceptually simple. Fast trapped-ion entangling gates have been preformed to-date only on two-ion registers \cite{Wong2017,Schafer2018}. 

Moreover, finding a power-efficient solution in the $N=2$ ions case and solving the quadratic problem for an all-to-all entangling gate in $N=3$ ions as well can be done in polynomial time, as shown in appendix \rom{5}. The $N=3$ ions solution is an excellent initial guess for numerically optimizing this problem for a larger number of ions.

In order to further justify omission of the carrier coupling term from Eq. \eqref{eqVI5}, beyond linear contributions, we use the second-order term of the Magnus expansion (see appendix \rom{4}). This generates additional quadratic constraints in $\boldsymbol{r}$, which correspond to two-photon processes that couple a qubit state to itself via side-band and carrier transitions. As shown numerically below, abiding these constraints is relatively easy.

We may reformulate the different constraints above as a constrained optimization problem. The resource we wish to optimize (minimize) is the field amplitude, as this is the relevant limit in terms of available laser power. Thus we form the problem,
\begin{equation}
    \underset{\left\{ \boldsymbol{x}\right\} }{\text{argmin}}\left(\left|K^T\boldsymbol{x}\right|_{1}\right)\text{ s.t }\begin{cases}
    \boldsymbol{x}^{T}\tilde{C}_{j}\boldsymbol{x}=\varphi_1^\text{desired}-\varphi_j^\text{desired}\\
    W\left(K^T\boldsymbol{x}\right)=0
    \end{cases}\label{eqOptProblem},
\end{equation}
where $W\left(\boldsymbol{r}\right)$ encapsulates the carrier-coupling quadratic constraints described above and $j=2,...,N$.

Note that in Eq. \eqref{eqOptProblem} we choose to minimize the 1-norm, i.e $\left|\boldsymbol{r}\right|_1=\sum_i\left|r_i\right|$. We are motivated by  $\Omega^2\left|\boldsymbol{r}\right|_1^2$ being the peak laser power during the gate. Furthermore, we are conceptually searching for generalized solutions of physically-motivated schemes which are in general spectrally sparse \cite{Sorensen2000,Palmero2017,Shapira2018,Webb2018}. We intend to violate this sparsity only weakly. The 1-norm favors solutions for which most entries of $\boldsymbol{r}$ are small.

\begin{figure}
\includegraphics[width=\columnwidth]{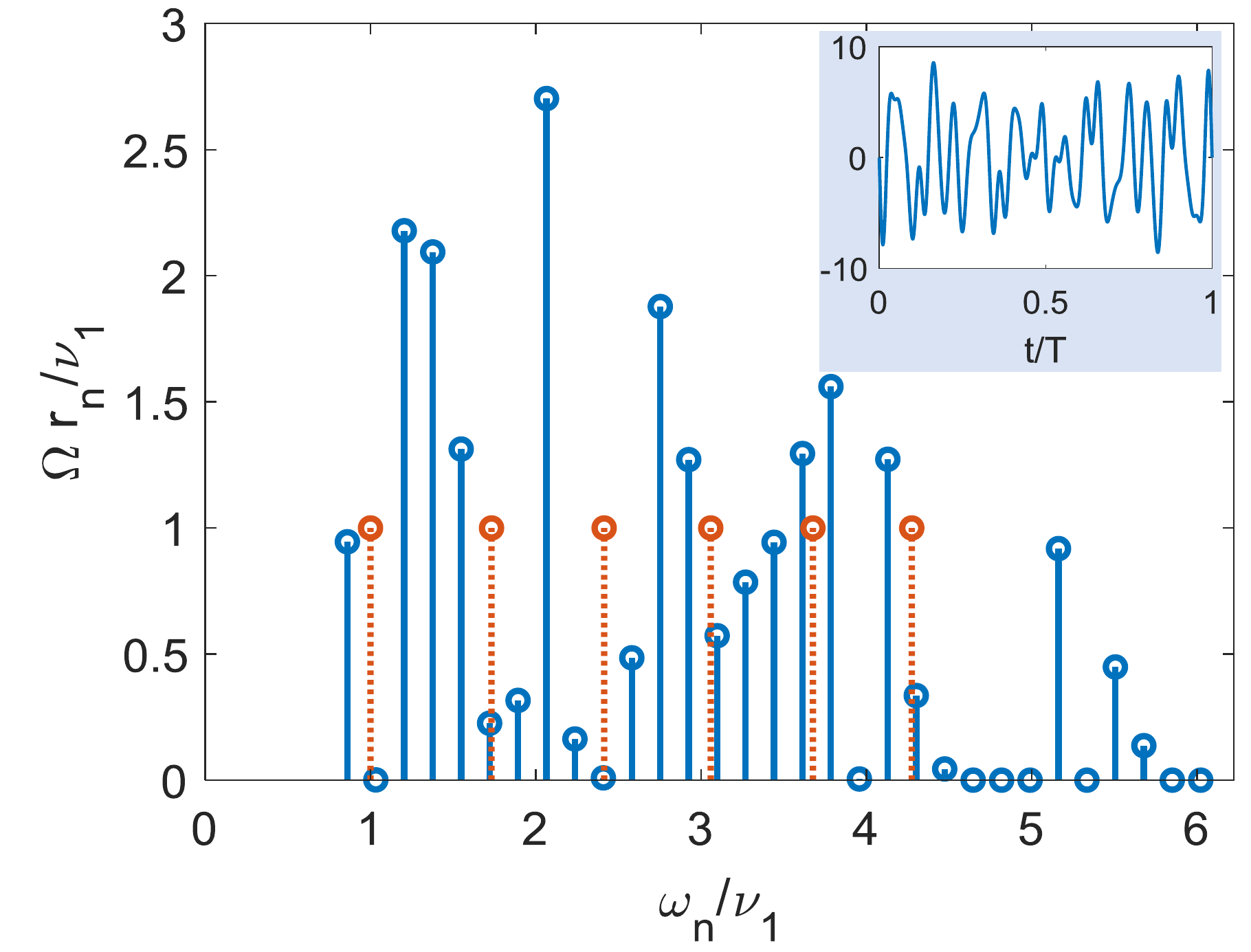}
\caption{Spectrum of highlighted example gate in Fig. \ref{figFidel6}. The results of the numerical search algorithm are the amplitudes of each driving harmonic (blue). The harmonics are centered around the normal mode frequencies (dashed red, height arbitrary). Together these generate a drive that abides all the constraints above. The inset shows the resulting pulse which starts and ends continuously at $0$.}\label{figSpectrum}\end{figure}

Figure \ref{figSpectrum} shows the required drive spectrum for the $N=6$ ions gate highlighted in Fig. \ref{figFidel6}. The drive is made of equally spaced tones, many of which have negligible amplitude due to the 1-norm optimization.

In order to obtain our entangling gates we use a constrained genetic numerical global search algorithm of Eq. \eqref{eqOptProblem}. The search algorithm outputs tone amplitudes, $\boldsymbol{r}$, from which we evaluate the resulting gate evolution and fidelity. We have arbitrarily set the tolerance of the constraints such that the resulting gate infidelity is lower than $10^{-4}$.

Our search algorithm is implemented using Matlab's global optimization toolbox and evaluated on a standard $3.6\text{ GHz}$ 8-core desktop computer. The algorithm runtime is determined by the number of degrees of freedom to optimize. Thus gates which operate at rates comparable to the trapping frequency are optimized faster than gate operating in the adiabatic regime.

\section{\rom{4}. Realization of all-to-all entanglement gates}

We present simulation results of all-to-all entangling gates. Our methods are valid for general trapped-ion architectures. For concreteness we focus here on trapped $^{88}\text{Sr}^+$ ions. We define the qubit states $\ket{0}\equiv\ket{5S_{\frac{1}{2},\text{-}\frac{1}{2}}}$ and $\ket{1}\equiv\ket{4D_{\frac{5}{2},\text{-}\frac{3}{2}}}$ as our qubit levels, which are coupled by an optical quadrupole transition at $674\text{ nm}$. We use the axial normal-modes of motion of a harmonic linear Paul trap, and take the frequency of the center-of-mass axial mode to be $400 \text{KHz}$. 

\begin{figure}
\includegraphics[width=\columnwidth]{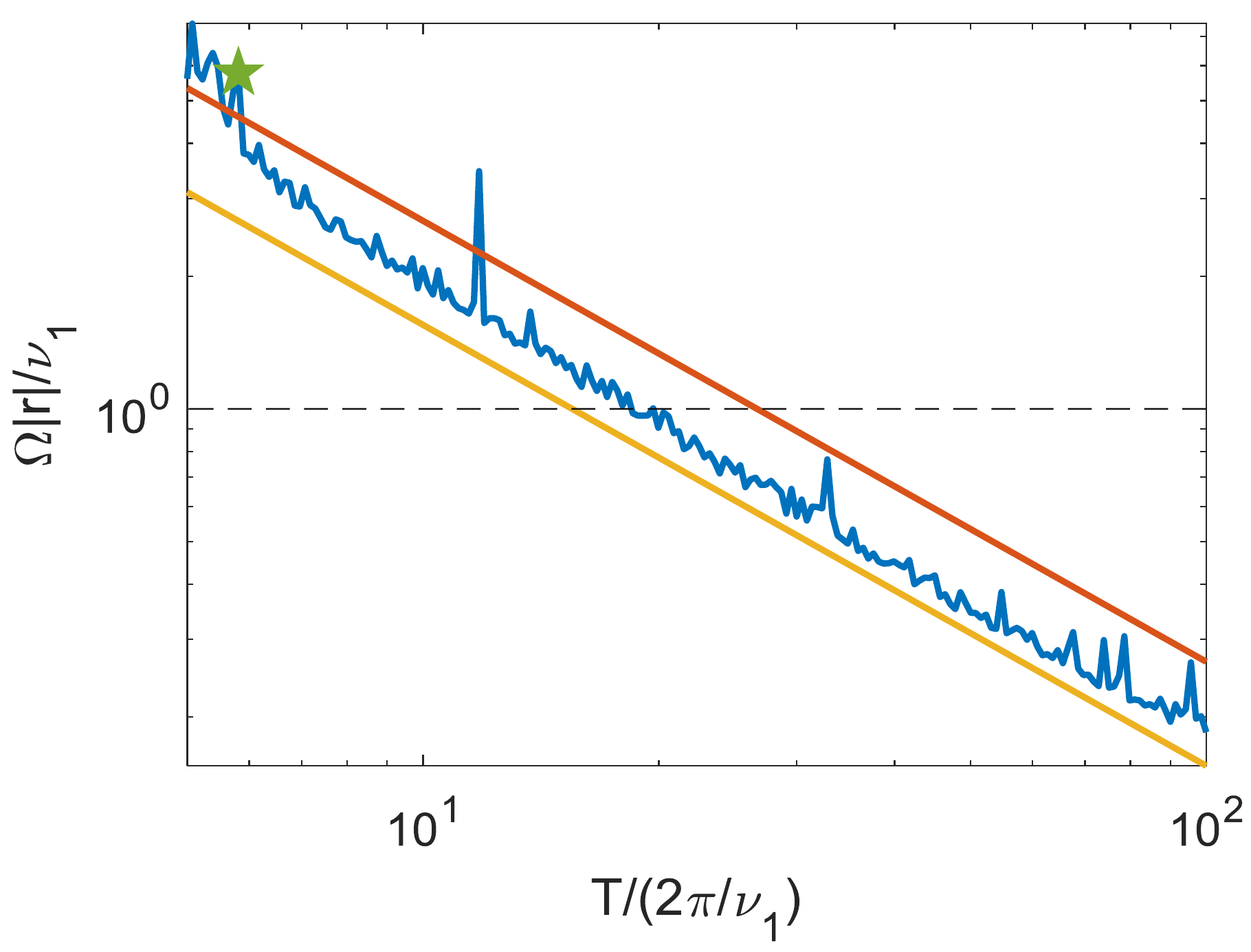}
\caption{Comparison of six ion multi-mode entangling gate drive amplitude (blue) to MS (yellow) and CarNu(2,3,7) (red) for varying gate times as in Fig. \ref{figFidel6}. The drive amplitude is measured in terms of $\nu_1$ (dashed black). The same example gate as in Fig. \ref{figFidel6} is highlighted (green star). All gates exhibit a similar scaling with respect to gate time. The overhead required to implement our gate is small, and starts to deviate only when the gate time approaches the secular trapping frequency}\label{figAmp6}\end{figure}

For an even number of ions we benchmark the performance of our all-to-all entangling gates via the fidelity of generating a GHZ state when acting on the ground state (for odd $N$ the resulting evolution does not generate GHZ states). This is sufficient as the GHZ states form a maximally sensitive set, which allows testing for coherent gate errors \cite{Gottesman2019}. The exact form of the fidelity is given in appendix \rom{6} and appendix \rom{7}.

In Fig. \ref{figFidel6} we show the resulting fidelity of different all-to-all entangling gates in a $N=6$ qubit register, with gate times between $100\frac{2\pi}{\nu_1}$ and $5\frac{2\pi}{\nu_1}$. As seen, the search algorithm finds solutions for which the infidelity is well below $10^{-4}$.

Figure \ref{figAmp6} shows the laser amplitude (or power; depending on the realisation), $\left|\boldsymbol{r}\right|\Omega$ in units of Rabi frequency, which is required for realizing our gates (blue), compared with the CarNu(2,3,7) gate (red) and MS gate (yellow). Clearly the required power is similar. The search algorithm runtime for gates with $T<20\frac{2\pi}{\nu_1}$ is approximately $5$ minutes.

\begin{figure}
\includegraphics[width=\columnwidth]{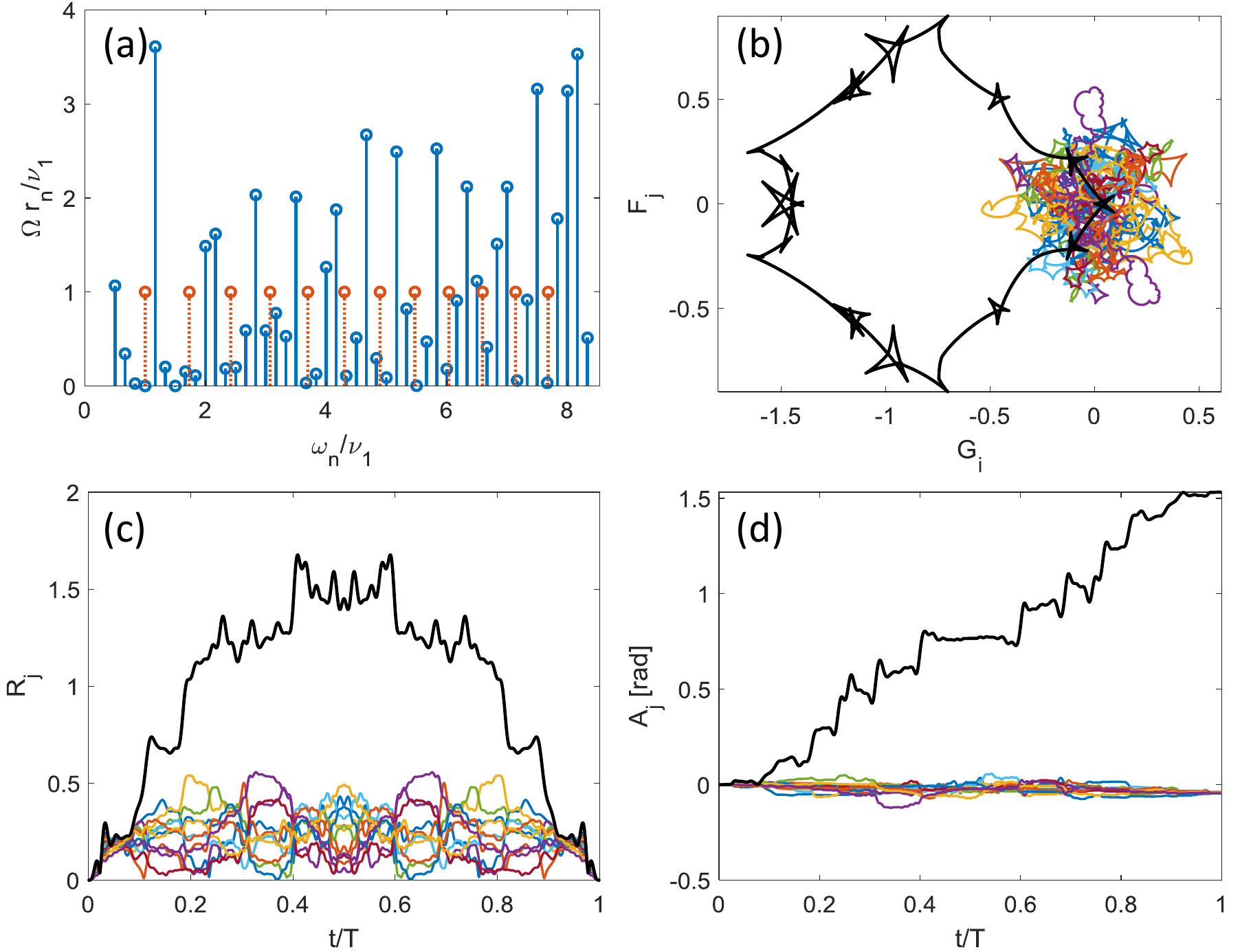}
\caption{Example entangling gate for $N=12$ ions. (a) Spectrum of laser drive. (b) Phase-space trajectory for center-of-mass mode (black) and remaining 11 modes (color). (c) Distance of phase-space trajectories from origin. Clearly all trajectories start and end at $0$. (d) Entangling phases for all modes. At gate time, the difference between the center of mass mode (black) and the remaining modes, which are all equal to each other, is approximately $\frac{\pi}{2}$, thus the fidelity of this gate is $F=0.9987$.}\label{figN12Example}\end{figure}

Figure \ref{figN12Example} shows a detailed analysis of a $N=12$ qubit gate, operating at $T=6\frac{2\pi}{\nu_1}$. Both linear and quadratic constraints are satisfied such that the resulting fidelity is $F=0.9987$, demonstrating that our method is applicable to larger qubit registers as well. In addition the gate is made robust to pulse timing errors, trapping frequency drifts and phonon-mode heating. The required laser power is $\left|\boldsymbol{r}\right|\Omega=10.26\nu_1$. The optimization algorithm runtime here is 105 minutes.

\section{\rom{5}. Realization of spin-Hamiltonians}

Our methods can also be used to generate spin-Hamiltonians for quantum simulations. We determine the required entanglement phases $\varphi_j^\text{ideal}$ that implement the unitary evolution operator $\exp\left(\sum_{i,k=1}^N j_{i,k}^\text{ideal}\hat{\sigma}_{y,i}\hat{\sigma}_{y,k}\right)$ at time $t=T$ and perform the same optimization described above.

The system state, after repeating the entanglement gate $n$ times, is equivalent to the evolution due to the Hamiltonian,
\begin{equation}
\hat{H}=\hbar\Omega \sum_{i,k=1}^N j_{i,k}^\text{ideal}\hat{\sigma}_{y,i}\hat{\sigma}_{y,k},\label{eqHamStrobo}    
\end{equation}
after an evolution time $t_n=\frac{n}{\Omega}$. This allows for a stroboscopic implementation of $\hat{H}$.

In addition, an effective Trotter-Hamiltonian of the form,
\begin{equation}
\hat{H}=\hbar\Omega \sum_{i,k=1}^N j_{i,k}\hat{\sigma}_{+,i}\hat{\sigma}_{-,k}+\text{H.c},\label{eqHamStroboTrot}   
\end{equation}
can be generated by interleaving $\hat{\sigma}_y$ and $\hat{\sigma}_x$ type interactions, which can be accomplished by a global $\frac{\pi}{2}$ phase shifts of the driving field.

In order to determine $\varphi_j^\text{ideal}$ we expand the desired coupling matrix, $j^\text{desired}$, in terms of the $\hat{J}_{j,y}^2$'s,
\begin{equation}
    \exp\left(i\sum_{i,k=1}^N j_{i,k}^\text{desired}\hat{\sigma}_{y,i}\hat{\sigma}_{y,k}\right)=\exp\left(i\sum_{j=1}^N \varphi_j\hat{J}_{y,j}^2\right).\label{eqJExpansion}
\end{equation}

Notably, the left-hand side of Eq. \eqref{eqJExpansion} has $\frac{1}{2}N\left(N+1\right)$ degrees of freedom and the right-hand side has only $N$ degrees of freedom, which means it cannot be generically solved. 

Equation \eqref{eqJExpansion} can be rewritten as the matrix equation,
\begin{equation}
    j^\text{desired}\cong\sum_{j=1}^N\varphi_j \boldsymbol{o}_j^T\boldsymbol{o}_j\label{eqJExpansionMat},
\end{equation}
with $\boldsymbol{o}_j$ the $j$'th row of $O$, i,e $(\boldsymbol{o}_j)_k=O_{j,k}$. The congruence symbol, $\cong$, defines a matrix equality up to the main diagonal, which is used here since the main diagonal contributes identity operators.

Equation \eqref{eqJExpansionMat} is linear in terms of the $\varphi_j$'s, and therefore is amenable to a least-squares approximation using the Moore–Penrose pseudoinverse method, yielding a solution $\varphi_j^\text{ideal}$ and the corresponding matrix $j^\text{ideal}=\sum_{j=1}^N\varphi_j^\text{ideal}\boldsymbol{o}_j^T\boldsymbol{o}_j$. 

The ideal implementation fidelity is then given by the normalized overlap,
\begin{equation}
F_\text{ideal}=\frac{1}{2}\left(1+\frac{\langle j^\text{ideal},j^\text{desired}\rangle}{\sqrt{\langle j^\text{ideal},j^\text{ideal}\rangle\langle j^\text{desired},j^\text{desired}\rangle}}\right)
\end{equation}
where we use the diagonal-less overlap $\langle A,B\rangle\equiv\sum_{n\neq m}A_{n,m}B_{m,n}$. 

We note that for higher spin-operators the congruence relation in Eq. \eqref{eqJExpansionMat} becomes an equality, which has a simpler solution, $\varphi_j=\boldsymbol{o}_j^T j \boldsymbol{o}_j$, and a lower ideal fidelity calculated with a trace inner-product.

We present simulation results of various spin-Hamiltonians. As in the section above we focus on trapped $^{88}\text{Sr}^+$ ions. Here we use the axial normal-modes of motion of an an-harmonic linear Paul trap designed such that the ions are equally spaced \cite{Johanning2016}. The frequency of the first axial mode is tuned to $400 \text{KHz}$.

\begin{figure}
\includegraphics[width=\columnwidth]{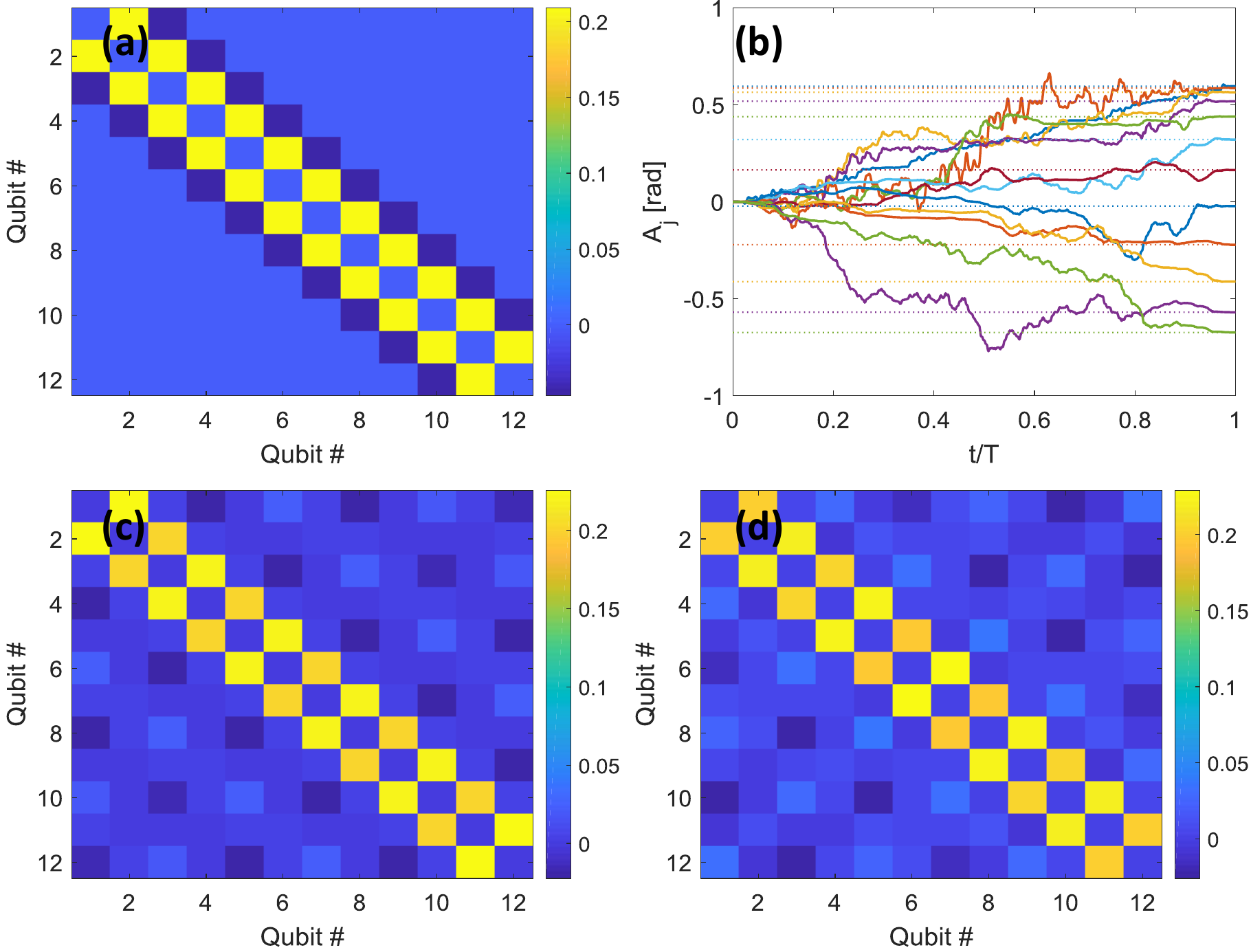}
\caption{More examples of simulated gates intended for analog quantum simulations. (a) Coupling matrix implementing nearest neighbour interaction with an opposite amplitude next-nearest neighbour, here we have set $j_{n,n+2}=-\frac{1}{4}j_{n,n+1}$ which is implemented with fidelity $F=0.9997$. (b) The resulting entanglement phase evolution of (a), clearly here each entanglement phase takes a distinct value at gate time (compared to Fig. \ref{figPhases} above). The corresponding desired entanglement phases are shown in dashed (c) Coupling matrix implementing the Su-Schrieffer-Heeger model in its trivial regime. The interlaced strong-weak pattern of the nearest neighbour couplings is easily seen. The fidelity is $F=0.9800$ (d) The same as in (c) in the topological non-trivial regime. The coupling pattern here is weak-strong. The implementation fidelity is $F=0.9758$.}\label{figMatMore}\end{figure}

In Fig. \ref{figMatNN} above we show the implemented coupling matrix of a nearest-neighbour model acting on a $N=12$ qubits, $j_{i,k}^{\text{n.n}}=\phi\left(\delta_{i,k+1}+\delta_{i,k-1}\right)$, for which the implementation fidelity is better than $0.999$ ($F_\text{ideal}=0.9999$). The gate time is $T=20\frac{2\pi}{\nu_1}$ and the required amplitude for $\phi=\frac{\pi}{4}$ is $\left|\boldsymbol{r}\right|\Omega=5.3\nu_1$.

Figure \ref{figMatMore} shows a small selection of more examples of possible simulation oriented entanglement gates for $N=12$ equally-spaced trapped-ion qubits, such as nearest-neighbours with opposite next-nearest-neighbours interaction coupling  (a), with fidelity of $F=0.9997$ ($F_\text{ideal}=0.9999$), corresponding to the Hamiltonian $\hat{H}=\hbar\Omega\sum_{n=1}^N\left(\hat{\sigma}_{y,n}\hat{\sigma}_{y,n+1}-\frac{1}{4}\sigma_{y,n}\sigma_{y,n+2}\right)$, and its resulting entanglement phase evolution (b), and the Su-Schriefer-Heeger model, i.e the coupling matrix $j_{i,k}^{\text{SSH}}=\left(\phi-(-1)^{\left(t+i+k\right)}\delta\phi\right)\left(\delta_{i,k+1}+\delta_{i,k-1}\right)$, such that $\phi>\delta\phi>0$, with $s=0$ in topological trivial regime (c) and $s=1$ in the non-trivial regime (d) \cite{Su1980}, with fidelity of $F=0.9800$ ($F_\text{ideal}=0.9801$) and $F=0.9758$ ($F_\text{ideal}=0.9768$) respectively. Clearly our method allows for the implementation of a variety of spin-Hamiltonians with close-to-ideal fidelities.

\section{\rom{6}. Conclusions}

We have presented a general method for designing multi-qubit entangling gates for trapped-ion qubits, implementing the evolution $\exp\left(\sum_{i,k=1}^N j_{i,k}\hat{\sigma}_{y,i}\hat{\sigma}_{y,k}\right)$. By utilizing all the normal-modes of motion of the ion-chain our gates operate outside of the adiabatic regime and can implement a variety of coupling matrices. Thus they may be used either as quantum-logic gates aimed at quantum computation, or in order to generate various spin-spin interactions for analog quantum simulations. 

Our gates require only a multi-tone global driving field, utilizing a bandwidth similar to that of the ion-chain's normal-modes. Our implementation results in a high-fidelity process, without a significant laser-amplitude overhead. Thus they are suited for many trapped-ion architectures. Furthermore, we have endowed our gates with robustness properties such that they are resilient to various noises and implementation imperfections.

\begin{acknowledgments}
This work was supported by the Israeli Science Foundation.
\end{acknowledgments}

\section{Appendix \rom{1}. Hamiltonian derivation}

We begin by deriving the Hamiltonian of $N$ trapped ions. The derivation follows at large Refs. \cite{Molmer11999,Sorensen2000}, however here we consider $N$ trapped ions and $N$ normal-modes of motion and do not use an adiabatic approximation with respect to the normal-mode frequencies.

The non-interacting lab-frame Hamiltonian is,
\begin{equation}
    \hat{H}_0=\sum_{k=1}^{N}\left(\hbar\nu_{k}\left(\hat{a}_{k}^{\dag}\hat{a}_{k}+\frac{1}{2}\right)+\frac{\hbar\omega_{0}}{2}\hat{\sigma}_{z,k}\right). \label{eqH02}
\end{equation}
with $\hat{a}_j$ the lowering operator of the $j$'th axial normal-mode of motion with frequency $\nu_j$, the single qubit separation frequency $\omega_0$ and $\hat{\sigma}_{z,k}$ the Pauli-$\hat{z}$ spin-operator acting on the $k$'th qubit.

The ions are driven by a multi-chromatic laser field, containing $2M$ frequencies arranged in pairs, $\left\{\omega_0\pm\omega_i\right\}_{i=1}^M$. Each component has phase $\phi_{\pm,i}=\pm\phi_i$, i.e the average phase of each pair is $0$, and each pair has the same amplitude $\Omega r_i$, with $\Omega$ a characteristic Rabi frequency and $r_i\in\mathbb{R}$ (such that $r_i\rightarrow -r_i$ is the same as $\phi_i\rightarrow\phi_i+\pi$). In total this driving field is determined by $3M$ degrees of freedom. 
The resulting interaction due to this field is,
\begin{equation}
    \hat{V}=2\hbar\Omega\sum_{i=1}^{M}r_{i}\sum_{n=1}^{N}\hat{\sigma}_{x,n}\cos\left(k\hat{x}_{n}-\omega_{0}t\right)\cos\left(\omega_{i}t+\phi_{i}\right),\label{eqV2}
\end{equation}
where $\hat{\sigma}_{x,n}$ is a Pauli-$\hat{x}$ spin-operator acting on the $n$'th qubit, $k$ is the laser momentum vector projected on the normal-mode direction of motion and $\hat{x}_n$ is the position operator of the $n$'th qubit. We note that we assumed implicitly that the ions are driven with a uniform global field, i.e $\Omega$ has no $n$-index. 

The driving applied on the qubits depends on the position of the ions, which has dynamics by itself, and thus this Hamiltonian couples the motion of the ions to the "spins" $\sigma$. 

The total Hamiltonian is, $\hat{H}=\hat{H}_0+\hat{V}$. We change to an interaction picture with respect to $\hat{H}_0$ to obtain,
\begin{equation}
\begin{split}
    \hat{V}_I & = \hbar\Omega\sum_{i=1}^{M}r_{i}\sum_{n=1}^{N}\cos\left(\omega_{i}t+\phi_{i}\right)\left(e^{-i\omega_{0}t}\hat{\sigma}_{+,n}+h.c\right) \\  & \cdot\left(e^{-i\sum\limits _{j=1}^{N}\eta_{j}O_{j,n}\left(\hat{a}_{j}^{\dagger}e^{i\nu_{j}t}+\hat{a}_{j}e^{-i\nu_{j}t}\right)-i\omega_{0}t}+h.c\right)
    ,\label{eqVI1}
\end{split}
\end{equation}
with $\hat{\sigma}_{+,n}$ the spin raising operator acting on the $n$'th ion, and $\eta_j\equiv k\sqrt{\frac{\hbar}{4\pi m\nu_j}}$, the Lamb-Dicke parameter of the $j$'th motional mode, with ion mass $m$. Furthermore, $O$ is an orthogonal matrix whose rows are the normal-modes of motion, such that the standard basis vectors are given by $\left(\textbf{e}_j\right)_i=\sum_{i=1}^{N}O_{i,j}$. The mode matrix $O$ can be determined in a semi-classical analysis \cite{James1998} and strongly depends on the effective trapping potential. Here we do not require specific knowledge of the normal-mode's structure, rather only that these orthogonal harmonic normal-modes exist. 

We note that the interaction in Eq. \eqref{eqVI1} contains counter-rotating terms at $\sim2\omega_0$, which is an optical frequency. These terms may be neglected in a rotating wave approximation (RWA). We obtain,
\begin{equation}
\begin{split}
    \hat{V}_{I} & =\hbar\Omega\sum_{i=1}^{M}r_{i}\cos\left(\omega_{i}t+\phi_{i}\right)  \\
    & \cdot\sum_{n=1}^{N}\left(e^{-i\sum\limits _{j=1}^{N}\eta_{j}O_{j,n}\left(\hat{a}_{j}^{\dagger}e^{i\nu_{j}t}+\hat{a}_{j}e^{-i\nu_{j}t}\right)}\hat{\sigma}_{+,n}+h.c\right).\label{eqVI2}
\end{split}
\end{equation}

Next we take the Lamb-Dicke approximation, i.e we assume that $\eta_j\ll1$ for all $j=1,...,N$ such that all normal-modes of motion are spectrally resolved. This simplifies the interaction in Eq. \eqref{eqVI2} further to,
\begin{equation}
\begin{split}
    \hat{V}_{I} & = \hbar\Omega\sum_{i=1}^{M}r_{i}\cos\left(\omega_{i}t+\phi_{i}\right)\sum_{n=1}^{N} \\ & \left[\left(\boldsymbol{1}-i\sum\limits _{j=1}^{N}\eta_{j}O_{j,n}\left(\hat{a}_{j}^{\dagger}e^{i\nu_{j}t}+\hat{a}_{j}e^{-i\nu_{j}t}\right)\right)\hat{\sigma}_{+,n}+h.c\right], \label{eqVI3}
\end{split}
\end{equation}
with quadratic corrections in $\eta_j$. The term proportional to $\boldsymbol{1}$ generates off-resonance carrier coupling. It is customary to neglect it in a RWA in terms of $\Omega\ll\nu_j$. Here however we intend not to perform such an adiabatic approximation. We nevertheless drop this term and justify it below by formulating constraints under which this term is effectively decoupled from the system's evolution. 

We are left with,
\begin{equation}
\begin{split}
    V_{I} & =\hbar\Omega\sum_{i=1}^{M}r_{i}\cos\left(\omega_{i}t+\phi_{i}\right) \\ & \cdot \sum_{j=1}^{N}\eta_{j}\left(\hat{a}_{j}^{\dagger}e^{i\nu_{j}t}+\hat{a}_{j}e^{-i\nu_{j}t}\right)\sum_{n=1}^{N}O_{j,n}\hat{\sigma}_{y,n}. \label{eqVI4}
\end{split}
\end{equation}

The three summations in Eq. \eqref{eqVI4} are on drive components, normal-modes an ions respectively. It is helpful to define the mode-dependent global Pauli spin operator as, $\hat{J}_{i,j}=\frac{\sqrt{N}}{2}\sum_{n=1}^{N}O_{j,n}\hat{\sigma}_{i,n}$, with $i\in\left\{x,y,z,+,-\right\}$ and $j\in\left\{1,...,N\right\}$. For simplicity we will assume the first normal-mode of motion is the center-of-mass mode, i.e $\hat{J}_{y,1}$ identifies with the global spin rotation $\hat{J}_y=\frac{1}{2}\sum_{n=1}^{N}\hat{\sigma}_{y,n}$.

Using this convention we are able to eliminate the latter summation on ions. Furthermore we define the normal-mode position (momentum) operator $\hat{q}_j=\frac{\hat{a}^\dagger_j+\hat{a}_j}{\sqrt{2}}$ ($\hat{p}_j=i\frac{\hat{a}^\dagger_j-\hat{a}_j}{\sqrt{2}}$), such that Eq. \eqref{eqVI4} becomes Eq. \eqref{eqVI5} of the main text.

\section{Appendix \rom{2}. Proof of sufficient entanglement phase constraint}
Here we prove the identity in Eq. \eqref{eqJSum} of the main text, i.e,
\begin{equation}
    \textbf{1}=e^{i\sum_{j=1}^N \hat{J}_{y,j}^2},\label{eqJSum2}
\end{equation}
with $\hat{J}_{i,j}=\frac{\sqrt{N}}{2}\sum_{n=1}^{N}O_{j,n}\hat{\sigma}_{i,n}$, such that $i\in{x,y,z,+,-}$ and $j=1,...,N$. We note that the columns of the mode-matrix, $O$, are orthonormal vectors.

Directly,
\begin{equation}
\begin{split}
    \sum_{j=1}^N \hat{J}_{y,j}^2 & = \frac{N}{4}\sum_{j,n,m=1}^N O_{j,n}O_{j,m}\hat{\sigma}_{y,n}\hat{\sigma}_{y,m} \\ &
    = \frac{N}{4}\sum_{j,n,m=1}^N O_{n,j}^TO_{j,m}\hat{\sigma}_{y,n}\hat{\sigma}_{y,m} \\ &
    = \frac{N}{4}\sum_{n,m=1}^N \delta_{n,m}\hat{\sigma}_{y,n}\hat{\sigma}_{y,m}  = \frac{N}{4}\sum_{n=1}^N \textbf{1}.
\end{split}\label{eqJSumProof}
\end{equation}
By exponentiation the first and last terms in Eq. \eqref{eqJSumProof} above we recover the identity up to an insignificant global phase.

We note that this result may be used not only to generate an all-to-all coupling via a center-of-mass mode, but also to generate any coupling scheme between the ions that can be written as a linear combination of the $\hat{J}_{y,j}^2$ operators, without the need to nullify contributions that do not appear in the explicit combination. 

For example, in order to generate a coupling of the form, $a\hat{J}_{y,1}^2+b\hat{J}_{y,2}^2$, instead of realizing $\varphi_1=a$, $\varphi_2=b$ and $\varphi_{j\geq3}=0$, which, due to the latter condition, is a hard task in the non-adiabatic regime, one may alternatively use $\varphi_1-\varphi_{j\geq3}=a$ and $\varphi_2-\varphi_{j\geq3}=b$, which is much less restrictive on all of the $j\geq3$ normal-modes.

\section{Appendix \rom{3}. Explicit expression for robustness properties}

As discussed in the main text, phase-space trajectory closure can be formulated as a linear constraint in the amplitudes vector, $\boldsymbol{r}$. Similarly, various robustness properties can as well be formulated as linear constraints.

Below we describe the matrix elements of $L$ which correspond to the different properties. The elements $L_{j,n}$ are given, which correspond to conditions applicable to the $j$'th normal mode and the $n$'th tone with frequency $\omega_n=\frac{2\pi}{T}n$. The desired property is obtained by satisfying the relation $\sum_{n=1}^N L_{j,n}r_n=0$ for all $j$.

Robustness to timing errors, i.e error of the form $T\rightarrow T+\delta T$, can be implemented by requiring that $\frac{dG_j}{d\delta T}|_{\delta T=0,\omega=\frac{2\pi}{T}n}=0$, and $\frac{dF_j}{d\delta T}|_{\delta T=0,\omega=\frac{2\pi}{T}n}=0$ \cite{Shapira2018}. We note that substitution of the harmonic frequencies should be done after differentiation (since the choice of frequencies does not depend on this kind of error). For a harmonic gate these terms vanish, for general frequencies we obtain,
\begin{equation}
\begin{split}
    L_{j,n}&=2\sqrt{2}\eta_j\cos\left(\nu_j T\right)\cos\left(\omega_n T+\phi_n\right)\\
    L_{j,n}&=2\sqrt{2}\eta_j\sin\left(\nu_j T\right)\cos\left(\omega_n T+\phi_n\right). \label{eqRobCar}
\end{split}
\end{equation}

We note that Eq. \eqref{eqRobCar} seemingly depends on the mode index $j$, however since we are only interested in the kernel of L, using $L_{j,n}=\cos\left(\omega_n T+\phi_n\right)$, suffices.

Higher-order robustness to timing errors may be easily implemented by requiring that higher-order derivatives vanish at the error-less gate time as well. All orders will remain linear in $\boldsymbol{r}$ and thus may be just as easily implemented.

Robustness to normal-mode errors, i.e errors of the form $\nu_j\rightarrow\nu_j+\delta\nu$, and normal-mode heating can similarly be minimized by requiring that $\int_{0}^{T}G_{j}\left(t\right)dt=0$, and $\int_{0}^{T}F_{j}\left(t\right)dt=0$. Which is easily seen by integration by parts of $\frac{d\alpha_j}{d\delta \nu}|_{\delta \nu=0}=0$. Similarly to robustnes to timing errors above, these constraints result in the matrix elements,
\begin{equation}
\begin{split}
    L_{j,n}&=\frac{2 \sqrt{2} \pi  n T^2 \eta _j \left(T \nu _j \left(2 \sin \left(T \nu _j\right)-T \nu _j\right)+4 \pi ^2 n^2\right)}{\left(T^2 \nu _j^2-4 \pi ^2 n^2\right){}^2}\\
    L_{j,n}&=-\frac{4 \sqrt{2} \pi  n T^3 \eta _j \nu _j \left(\cos \left(T \nu _j\right)-1\right)}{\left(T^2 \nu _j^2-4 \pi ^2 n^2\right){}^2}. \label{eqRobNu}
\end{split}
\end{equation}

In Raman gate configurations a possible source of error is a phase drift between the two counter-propagating Raman beams (in direct-transition gates this corresponds to phase noise in the RF signal generators and is less likely). Robustness to this error can be obtained with the matrix elements,
\begin{equation}
\begin{split}
    L_{j,n}&=\frac{\sqrt{2} T^2 \eta _j \nu _j \sin \left(T \nu _j\right)}{T^2 \nu _j^2-4 \pi ^2 n^2}\\
    L_{j,n}&=\frac{\sqrt{2} T^2 \eta _j \nu _j \left(\cos \left(T \nu _j\right)-1\right)}{4 \pi ^2 n^2-T^2 \nu _j^2}. \label{eqRobOpt}
\end{split}
\end{equation}

\section{Appendix \rom{4}. Magnus expansion for carrier coupling}
As mentioned above, we justify the omission of the carrier coupling term in the derivation of Eq. \eqref{eqVI4} by nulling the term's contributions in a Magnus expansion. Specifically, we rewrite the Hamiltonian in Eq. \eqref{eqVI3} as the sum of the non-commuting terms, $\hat{V}_I=\hat{H}_{c.c}+\hat{H}_{MS}$, with $\hat{H}_{c.c}=\hbar\Omega\sum_{i=1}^{M}r_{i}\cos\left(\omega_{i}t+\phi_{i}\right)\hat{J}_{x,1}$ and $\hat{H}_{MS}$ is given by Eq. \eqref{eqVI5}.

Following Ref. \cite{Blanes2009} we expand the unitary evolution operator $\hat{U}$, due to $\hat{V}_I$, to second order,
\begin{equation}
\begin{split}
      \hat{U} & =\exp\left(\sum_k \hat{\Omega}_k\right)\\
      \hat{\Omega}_1 & =-\frac{i}{\hbar}\int_0^T dt_1 \hat{V}_I\left(t_1\right) \\
      \hat{\Omega}_2 & =\frac{1}{2}\left(-\frac{i}{\hbar}\right)^{2}\int\limits _{0}^{t}dt_{1}\int\limits _{0}^{t_{1}}dt_{2}\left[\hat{V}_I\left(t_1\right),\hat{V}_I\left(t_2\right)\right].
\end{split}\label{eqMagnus}
\end{equation}

In the first order we obtain,
\begin{equation}
    \hat{\Omega}_1=-i\Omega\sum_{n=1}^{M}r_{n}\int\limits _{0}^{T}dt\sin\left(\omega_{n}t\right)\hat{J}_{x,1}+\hat{\Omega}_{1,MS},\label{eqOm1}
\end{equation}
where $\hat{\Omega}_{1,MS}$ corresponds to desired terms that are not due to carrier coupling (these create displacement). We note that the first term in Eq. \eqref{eqOm1} vanishes identically in the harmonic basis (however does not vanish in the conventional MS gate).

In the second order we again obtain desired terms that are not due to carrier coupling (generating the entanglement phases) and carrier coupling related terms. These terms are,
\begin{equation}
\begin{split}
    \hat{\Omega}_{2,cc} & =  -\frac{i\Omega^2}{4}\sum_{n,m=1}^{M}r_n r_m\int_0^T dt_1 \bigg[ \sin\left(\omega_n T\right)\sum_{j=1}^N \eta_j \\
    & \cdot \int_0^{t_1} dt_2 \sin\left(\omega_m t_2\right) \left(\cos\left(\nu_j t_2\right)\hat{p}_j+\sin\left(\nu_j t_2\right)\hat{q}_j\right)\hat{J}_{z,j}\bigg].\label{eqOm2}
\end{split}
\end{equation}

The evolution due to $\hat{\Omega}_{2,cc}$ corresponds to two-photon processes involving a side-band transition and a carrier transition, generating a mode-dependent effective energy shift of the qubit levels due to the $\hat{J}_{z,j}$ operator. 

Furthermore, similarly to what we have seen in the "normal" gate evolution, in Eq. \eqref{eqU}, the evolution can be pictured along phase-space trajectories, $\left(G_{j,cc}\left(t\right),F_{j,cc}\left(t\right)\right)$, where $G_{j,cc}$ ($F_{j,cc}$) is the term proportional to $\hat{p}_j$ ($\hat{x}_j$). Since, in general, the trajectories do not close at $t=T$ an additional infidelity penalty occurs due to residual entanglement to the motional degrees of freedom.

\section{Appendix \rom{5}. Explicit solutions of the quadratic constraints for $N=2,3$}

As we stated in the main text, the quadratic constraint in Eq. \eqref{eqAmpCons} is an NP-hard problem. Here we show that for the $N=2$ ions it is easy to construct solutions that are power efficient and that for the $N=3$ ions it is easy to construct solutions, however their efficiency is a-priori unknown.

As shown above, by restricting the quadratic problem to the kernel of the linear constraints matrix $L$, we have reduced the entangling gate problem to satisfying the $N-1$ quadratic equations $\boldsymbol{x}^T \tilde{C}_j\boldsymbol{x}=\frac{\pi}{2}$, for $j=2,...,N$, where $\boldsymbol{x}$ is an $l$-element real vector and $l$ is the dimension of the kernel of $L$ (the number of independent solutions to the linear constraints). An efficient solution is a solution which satisfies the $N-1$ equations while minimizing $\left|\boldsymbol{r}\right|$.

For the $N=2$ there is a single quadratic equation, $\boldsymbol{x}^T \tilde{C}_2\boldsymbol{x}=\frac{\pi}{2}$. Any arbitrary vector $\boldsymbol{x}$ satisfies by-definition the linear constraint and takes some value, $C=\boldsymbol{x}^T \tilde{C}_2\boldsymbol{x}$. By renormalizing $\boldsymbol{x}\rightarrow \sqrt{\frac{\pi}{2\left|C\right|}}\boldsymbol{x}$, we obtain a valid solution. We note that if $C<0$ we actually generate the entangling phase $-\frac{\pi}{2}$ which also generates a GHZ state.

As $\boldsymbol{x}$ is arbitrary, this method does not ensure the solution efficiency. In order to obtain an efficient solution we note that $\tilde{C}_2$ is symmetric and therefore can be diagonalized. Every eigenvector of it, which corresponds to a positive eigenvalue, can be a solution. Specifically, the eigenvector corresponding to the largest eigenvalue will be the optimal solution, i.e $\boldsymbol{x}=\boldsymbol{u}_1\sqrt{\frac{\pi}{2\lambda_1}}$, where $\boldsymbol{u}_1$ is a normalized eigenvector of $\tilde{C}_2$, corresponding to the largest eigenvalue $\lambda_1>0$. As above, if there are no positive eigenvalues we may pick the largest eigenvalue in absolute value and generate a $-\frac{\pi}{2}$ phase.

We now proceed to the $N=3$ solution. We define $\tilde{D}_j=\tilde{C}_2-\tilde{C}_j$ with $j=3,4,...,N$, which are also symmetric real $l\times l$ matrices. The quadratic constraint above becomes,
\begin{equation}
\begin{cases}
    \boldsymbol{x}^T\tilde{D}_j\boldsymbol{x}=0 & j=3,...,N\\
    \boldsymbol{x}^T\tilde{C}_2\boldsymbol{x}=\pm\frac{\pi}{2}.
\end{cases}\label{eqAmpConsD}
\end{equation}

For $N=3$ we only have one of these matrices, $D_3$, which can be spectrally decomposed to,
\begin{equation}
    \tilde{D}_3=\sum_{i=1}^p \lambda_i\boldsymbol{\psi}_i\boldsymbol{\psi}_i^T+\sum_{j=1}^n \gamma_j\boldsymbol{\xi}_j\boldsymbol{\xi}_j^T, \label{eqD3}
\end{equation}

where the $\boldsymbol{\psi}$'s and $\boldsymbol{\xi}$'s are normalized eigenvectors of $\tilde{D}_3$ corresponding to the positive eigenvalues $\lambda_i$ with $i=1,...,p$ and negative eigenvalues $\gamma_j$ with $j=1,...,n$, respectively. 

Assuming that $n,p>0$, i.e that $\tilde{D}_3$ has both positive and negative eignavlues, we choose an arbitrary positive eigenvalue and negative eigenvalue and set,
\begin{equation}
    \boldsymbol{x} =C\left(\boldsymbol{\psi}_i+\sqrt{\left|\frac{\lambda_i}{\gamma_j}\right|}\right)\text{  }i\in\left\{1,...,p\right\}\text{  }j\in\left\{1,...,n\right\}.\label{eqxforN3}
\end{equation}
This choice suffices such that $\boldsymbol{x}^T\tilde{C}_2\boldsymbol{x}=\boldsymbol{x}^T\tilde{C}_3\boldsymbol{x}$. The normalization $C$ is chosen such that $\boldsymbol{x}^T\tilde{C}_2\boldsymbol{x}=\pm\frac{\pi}{2}$, thus satisfying Eq. \eqref{eqAmpConsD}. This solution can fail if the resulting $\boldsymbol{x}$ is an eigenvector of one of the $\tilde{C}_j$'s with a zero eigenvalue, however this is not generic.

We note that if the eigenvalues of any of the $\tilde{D}_j$'s are only positive or only negative then the problem cannot be solved.

The solution for $N=3$ above implies a general approach for numerically searching for a solution for an arbitrary number of ions. In each step of the nuermical search a candidate $\boldsymbol{x}$ is evaluated for feasibility, i.e whether it satisfies the quadratic constraints, and optimiality, i.e whether it corresponds to a low-power solution.

We may improve upon the candidate $\boldsymbol{x}$ by renormalizing it such that it at least satisfies $\boldsymbol{x}^T\tilde{D}_3\boldsymbol{x}=0$. This is done by expanding $\boldsymbol{x}$ with the positive and negative sub-spaces of $\tilde{D}_3$, that is,
\begin{equation}
    \boldsymbol{x}=\sum_{i=1}^p a_i \boldsymbol{\psi}_i+\sum_{j=1}^n b_j \boldsymbol{\xi}_j.\label{eqxProjected}
\end{equation}

Such that $\boldsymbol{x}^T\tilde{D}_3\boldsymbol{x}=\sum_{i=1}^p a_i^2\lambda_i+\sum_{j=1}^n b_j^2\gamma_j$. We note that in order for this expression to vanish the positive sum must be equal to the magnitude of the negative sum. 

Thus we define the vectors $\tilde{\boldsymbol{a}}$ ($\tilde{\boldsymbol{b}}$), with the elements $\tilde{a}_i=\frac{a_i}{\sqrt{\lambda_i}}$ ($\tilde{b}_j=\frac{b_j}{\sqrt{\left|\gamma_j\right|}}$). By renormalizing $\tilde{\boldsymbol{a}}\rightarrow\tilde{\boldsymbol{a}}/\left|\tilde{\boldsymbol{a}}\right|$ ($\tilde{\boldsymbol{b}}\rightarrow\tilde{\boldsymbol{b}}/\left|\tilde{\boldsymbol{a}}\right|$), i.e such that  they lie on the $p$-dimensional and $n$-dimensional unit spheres respectively then $\boldsymbol{x}^T\tilde{D}_3\boldsymbol{x}=0$ is satisfied. Finally, we renormalize the resulting $\boldsymbol{x}$ such that $\boldsymbol{x}^T\tilde{C}_2\boldsymbol{x}=\frac{\pi}{2}$ and Eq. \eqref{eqAmpConsD} is satisfied.

We note that the solutions presented here treats the $\tilde{C}_j$'s as arbitrary. The numerical solution can possibly be sped-up by taking advantage of the problem's underlying structure, i.e that the matrices originate from the contributions of different harmonics to the entanglement phases.

\section{Appendix \rom{6}. Unitary Fidelity calculations}

We separate the all-to-all gate fidelity to two contributions, unitary fidelity, $F_U$, which is determined by deviations of the state functions, $\left\{G_j,F_j,A_j\right\}_{j=1}^N$ from their ideal values at the gate time, and carrier-coupling fidelity, $F_\text{c.c}$, which is determined by the effect of the carrier-coupling Hamiltonian, $H_{cc}$, described above. Assuming both errors are small then we calculate the total gate infidelity as,
\begin{equation}
    I_\text{total}=1-F_\text{total}\approx1-F_U F_\text{c.c}.\label{eqInfidel}
\end{equation}

Here we derive expressions for $F_U$. Derivation of $F_{c.c}$ appears in appendix \rom{7} below. Throughout our derivations we assume that $N$ is even.

We define, $F_U=\broket{GHZ}{\hat{\rho}_q\left(T\right)}{GHZ}$, with $\hat{\rho}_q\left(t\right)$ the qubit-subspace density matrix, after evolution time $t$.

In order to avoid direct evolution of the state in a $\left(2^N\cdot n_\text{max}^N\right)$-dimensional Hilbert space, with $n_\text{max}$ the maximum phonon number of the different normal-modes, we first obtain a more efficient expression. 

Following a similar derivation as in \cite{Roos2008}, we note the identity,
\begin{equation}
\begin{split}
    \hat{U}_{j}	& =e^{-iA_{j}\hat{J}_{y,j}^{2}}e^{-iF_{j}\hat{x}_{j}\hat{J}_{y,j}}e^{-iG_{j}\hat{p}_{j}\hat{J}_{y,j}} \\ &
	=e^{-i\left(A_{j}+\frac{F_{j}G_{j}}{2}\right)J_{y,j}^{2}}e^{-i\left(G_{j}\hat{p}_{j}+F_{j}\hat{x}_{j}\right)J_{y,j}} \\ &
	=e^{-i\left(A_{j}+\frac{F_{j}G_{j}}{2}\right)\hat{J}_{y,j}^{2}}\hat{D}\left(\alpha_{j}\hat{J}_{y,j}\right)
\end{split},\label{eqUtoD}
\end{equation}
where for brevity we omit the time-dependence of $G_j$ , $F_j$ and $A_j$, and used the displacement operator, $\hat{D}_{j}\left(\alpha\right)=\exp\left(\alpha\hat{a}_{j}^{\dag}-\alpha^{\ast}\hat{a}_{j}\right)$, such that here $\alpha_{j}=-\frac{i}{\sqrt{2}}\left(F_{j}+iG_{j}\right)$.

Furthermore, we note that $\hat{D}_j\left(\alpha\hat{J}_{y,j}\right)=\sum_i \hat{D}_j\left(\alpha_j\lambda_{j,i}\right)\hat{P}_{j,i}\lambda_{j,i}$, where $\hat{P}_{j,i}$ is a projector to the subspace spanned by the $i$'th eigenvector of $\hat{J}_{y,j}$, with eigenvalue $\lambda_{j,i}$.

This allows us to rewrite the evolution operator in Eq. \eqref{eqU} as,
\begin{equation}
\begin{split}
\hat{U} & =\prod_{j}\left(e^{-i\left(A_{j}+\frac{F_{j}G_{j}}{2}\right)\hat{J}_{y,j}^{2}}\sum_{i}P_{i}D_{j}\left(\alpha_{j}\lambda_{j,i}\right)\right) \\ & 
=\prod_{j}\left(\sum_{i}\hat{Q}_{j,i}\hat{D}_{j}\left(\alpha_{j}\lambda_{j,i}\right)\right),
\end{split}\label{eqU2}
\end{equation}
where the operator, $\hat{Q}_{j,i}=\hat{P}_{i}e^{-i\left(A_{j}+\frac{F_{j}G_{j}}{2}\right)\lambda_{j,i}^{2}}$ acts exclusively in the qubit subspace. We note that we dropped the mode-index $j$ from the projector $\hat{P}_i$ as all the $\hat{J}_{y,j}$ operators have the same eigenvectors (and differ only by eigenvalues).

Using the form of $\hat{U}$ in Eq. \eqref{eqU2} above we able to easily trace out the normal-mode degrees of freedom. We have,
\begin{equation}
\begin{split}
    \hat{\rho}_q&=\sum_{\boldsymbol{n}}\broket{\boldsymbol{n}}{\hat{U}\hat{\rho}_0\hat{U}^\dagger}{\boldsymbol{n}} \\ &
    = \sum_{\alpha,\beta}\bigg[\hat{P}_\alpha\left(\prod_{j_1}\hat{Q}_{j_1,\alpha}\right)\hat{\rho}_{q,0}\left(\prod_{j_2}\hat{Q}_{j_2,\beta}\right)\hat{P}_\beta \\ &
    \cdot \prod_j\sum_{n_j}\broket{n_j}{\hat{D}_j\left(\alpha_j \lambda_{j,\alpha}\right)\hat{\rho}_{j,0}\hat{D}_j\left(\alpha_j \lambda_{j,\beta}\right)}{n_j}\bigg],
\end{split}\label{eqTraceOut}
\end{equation}
where $\hat{\rho}_0=\hat{\rho}_{q,0}\otimes\hat{\rho}_{1,0}\otimes\cdots\otimes\hat{\rho}_{N,0}$ is the system initial states, assumed to be made of the qubit ground state $\hat{\rho}_{q,0}$ and normal-mode thermal states, $\hat{\rho}_{j,0}$, with $j=1,...,N$, such that the probability of the $n$'th phonon state is $p_{n}=\frac{1}{\bar{n}_j+1}\left(\frac{\bar{n}_j}{\bar{n}_j+1}\right)^{n}$, where $\bar{n}_j$ is the average occupation number of the $j$'th normal-mode.

To proceed we use the identity, relevant to thermal states \cite{Roos2008},
\begin{equation}
\sum_{n}\broket n{\hat{D}\left(\alpha\lambda_{\alpha}\right)\hat{\rho}_{j,0}\hat{D}\left(\alpha\lambda_{\beta}\right)}{n}=e^{-\left|\alpha\right|^{2}\left(\lambda_{\alpha}-\lambda_{\beta}\right)\left(\bar{n}+\frac{1}{2}\right)}.\label{eqThermIdent}
\end{equation}

Thus we obtain,
\begin{equation}
\begin{split}
    \hat{\rho}_q & =\sum_{\alpha,\beta}\bigg[P_{\alpha}\rho_{q,0}P_{\beta} \\ &
    \cdot\prod_{j}e^{-i\left(A_{j}+\frac{F_{j}G_{j}}{2}\right)\left(\lambda_{j,\alpha}^{2}-\lambda_{j,\beta}^{2}\right)}e^{-\frac{R_{j}^{2}}{2}\left(\lambda_{j,\alpha}-\lambda_{j,\beta}\right)^{2}\left(\bar{n}_{j}+\frac{1}{2}\right)}\bigg],
\end{split}\label{eqTraceOut2}
\end{equation}
with $R_j=G_j^2+F_j^2$.

Since the qubit ground state, written in the $\hat{J}_{y,1}$ basis, is an equal superposition of all states, then in this basis Eq. \eqref{eqTraceOut2} becomes,
\begin{equation}
\begin{split}
    \hat{\rho}_q & =\sum_{\alpha,\beta}\bigg[\ketbra{\alpha}{\beta} \\ &
    \cdot\prod_{j}e^{-i\left(A_{j}+\frac{F_{j}G_{j}}{2}\right)\left(\lambda_{j,\alpha}^{2}-\lambda_{j,\beta}^{2}\right)}e^{-\frac{R_{j}^{2}}{2}\left(\lambda_{j,\alpha}-\lambda_{j,\beta}\right)^{2}\left(\bar{n}_{j}+\frac{1}{2}\right)}\bigg].
\end{split}\label{eqTraceOut3}
\end{equation}

A simple way to calculate $F_U$ is by computing, $F_U=\text{Tr}\left[\hat{\rho}_q\hat{\rho}_\text{GHZ}\right]$, 
where $\hat{\rho}_\text{GHZ}$ is obtained by setting $G_j=F_j=0$, $A_1=\frac{\pi}{2}$ and $A_{j\geq2}=0$ in Eq. \eqref{eqTraceOut3} above.

Alternatively in this basis the GHZ state can be written as,
\begin{equation}
    \ket{GHZ}=\frac{1}{\sqrt{2^{N+1}}}\sum_{\alpha}\left(1-i\left(-1\right)^{\frac{N}{2}}P\left(\alpha\right)\right)\ket{\alpha},\label{eqGHZ}
\end{equation}
where $P\left(\alpha\right)$ is the state parity, i.e it takes the value 1 if there are an even number of qubits in the state $\ket{+i}$ and $-1$ otherwise, and we have assumed that $N$ is even. Thus the unitary fidelity is explicitly given by,
\begin{equation}
\begin{split}
    F_U & = \frac{1}{2^{2N+1}}\sum_{\alpha,\beta}\bigg[\left(1+i\left(-1\right)^{\frac{N}{2}}P\left(\alpha\right)\right)\left(1-i\left(-1\right)^{\frac{N}{2}}P\left(\beta\right)\right) \\ &
    \cdot \prod_{j=1}^{N}e^{-i\left(A_{j}+\frac{F_{j}G_{j}}{2}\right)\left(\lambda_{j,\alpha}^{2}-\lambda_{j,\beta}^{2}\right)}e^{-\frac{R_{j}^{2}}{2}\left(\lambda_{j,\alpha}-\lambda_{j,\beta}\right)^{2}\left(\bar{n}_{j}+\frac{1}{2}\right)}\bigg]. \label{eqFU}
\end{split}
\end{equation}

We note that the expression in Eqs. \eqref{eqTraceOut3} and \eqref{eqFU} use a double summation on $N$-qubit states, thus their evaluation requires $\mathcal{O}\left(2^{2N}\right)$ calculations. 

We note that if we are coupled exclusively to the center-of-mass mode, we can reduce the number of calculations by exploiting the structure of the eigenvalues of $\hat{J}_{y,1}$. Namely instead of summing on states, as in Eq. \eqref{eqTraceOut3}, we sum on the eigenvalues $-\frac{N}{2},-\frac{N}{2}+1,...,\frac{N}{2}$. We get,

\begin{equation}
\begin{split}
    F_{U,1} & = \frac{1}{2^{2N+1}}\sum_{\lambda_{\alpha},\lambda_{\beta}=-\frac{N}{2}}^{\frac{N}{2}}\bigg[{N \choose \frac{N}{2}+\lambda_{\alpha}}{N \choose\frac{N}{2}+\lambda_{\beta}} \\ & 
    \cdot \left(1-i\left(-1\right)^{\lambda_{\alpha}}\right)\left(1+i\left(-1\right)^{\lambda_{\beta}}\right)e^{-i\left(A+\frac{FG}{2}\right)\left(\lambda_{\alpha}^{2}-\lambda_{\beta}^{2}\right)} \\ &
    \cdot e^{-\frac{F^{2}+G^{2}}{2}\left(\lambda_{\alpha}-\lambda_{\beta}\right)^{2}\left(\bar{n}+\frac{1}{2}\right)}\bigg]. \label{eqFU1}
\end{split}
\end{equation}

This expression can be evaluated with $\mathcal{O}\left(N^2\right)$ calculations. 

Similarly, the fidelity of remaining in the ground state when coupled exclusively to the center-of-mass mode, is given by,
\begin{equation}
\begin{split}
    F_{I,1} & = \bigg[ \frac{1}{2^{2N}}\sum_{\lambda_{\alpha},\lambda_{\beta}=-\frac{N}{2}}^{\frac{N}{2}}{N \choose \frac{N}{2}+\lambda_{\alpha}}{N \choose \frac{N}{2}+\lambda_{\beta}} \\ & \cdot 
    e^{-i\left(A+\frac{FG}{2}\right)\left(\lambda_{\alpha}^{2}-\lambda_{\beta}^{2}\right)}e^{-\frac{F^{2}+G^{2}}{2}\left(\lambda_{\alpha}-\lambda_{\beta}\right)^{2}\left(\bar{n}+\frac{1}{2}\right)}\bigg].\label{eqFI1}    
\end{split}
\end{equation}

Utilizing the entanglement phase identity in Eq. \eqref{eqJSum} we obtain a simple approximation for $F_U$,
\begin{equation}
    F_U\approx F_{U,1}\left(A_1-\bar{A},G_1,F_1\right)\prod_{j=2}^N F_{I,1}\left(A_j-\bar{A},G_j,F_j\right),
\end{equation}
with $\bar{A}=\frac{1}{N-1}\sum_{n=2}^N A_n$. That is, we use the center-of-mass fidelity in Eq. \eqref{eqFU1}, with the mean difference between $A_1$ and the other entanglement phases, and the identity center-of-mass fidelity in Eq. \eqref{eqFI1} to calculate the "excess" phase. Using this expression $F_U$ may be approximated with $\mathcal{O}\left(N^3\right)$ calculations. 

Nevertheless, in the simulations presented in the main text we use the full expression for $F_U$.

\section{Appendix \rom{7}. Carrier Coupling Fidelity Calculations}

As mentioned in appendix \rom{4}, for non-harmonic gates, the first order Magnus contribution of the carrier coupling terms does not vanish. The infidelity due to these terms has been previously evaluated as \cite{Shapira2018},
\begin{equation}
    F_{cc,1}=\cos\left(2\Omega\sum_{i=1}^M r_i\frac{\cos\left(\omega_i T+\phi_i\right)}{\omega_i}\right).\label{eqFcc1}
\end{equation}

Furthermore, the second order Magnus terms, derived in \rom{4}, contribute to the carrier-coupling infidelity since the trajectories formed by them, $\left(G_{j,cc}\left(t\right),F_{j,cc}\left(t\right)\right)$, do not generally close and thus leave the spin and motional degrees of freedom entangled.

In analogy to the derivation of the unitary fidelity in appendix \rom{6} we may calculate the resulting trajectory formed by these terms and evaluate the resulting carrier coupling infidelity. 

For simplicity we use thw two-ion fidelity analogue,
\begin{equation}
\begin{split}
    F_{cc,2}& =\prod_{j=1}^{N}\bigg[\frac{3+e^{-\left(F_{j,cc}^{2}+G_{j,cc}^{2}\right)}}{8} \\ & +\frac{1}{2}\cos\left(\frac{F_{j,cc}G_{j,cc}}{2}\right)e^{-\frac{F_{j,cc}^{2}+G_{j,cc}^{2}}{4}}\bigg], \label{eqFcc2}
\end{split}
\end{equation}
that is, we use the 2-qubit identity fidelity assuming all modes are a center-of-mass mode. Finally,  $F_{cc}=F_{cc,1}F_{cc,2}$.


\begin{thebibliography}{99}

\bibitem{DiVincenzo1995}
D. P. DiVincenzo, Two-bit gates are universal for quantum computation, Physical Review A 51, 1015 (1995).

\bibitem{Barenco1995}
A. Barenco, C. H. Bennett, R. Cleve, D. P. DiVincenzo, N. Margolus, P. Shor, T. Sleator, J. A. Smolin, and H. Weinfurter, Elementary gates for quantum computation, Physical Review A, 52 3457 (1995).

\bibitem{Kitaev1997}
A. Y. Kitaev, Quantum computations: algorithms and error correction, Russian Mathematical Survey 52(6), 1191-1249 (1997).

\bibitem{Myerson2008}
A. H. Myerson, D. J. Szwer, S. C. Webster, D. T. C. Allcock, M. J. Curtis, G. Imreh, J. A. Sherman, D. N. Stacey, A. M. Steane, and D. M. Lucas, High-Fidelity Readout of Trapped-Ion Qubits, Physical Review Letters 100, 200502 (2008).

\bibitem{Harty2014}
T. P. Harty, D. T. C. Allcock, C. J. Ballance, L. Guidoni, H. A. Janacek, N. M. Linke, D. N. Stacey, and D. M. Lucas, High-Fidelity Preparation, Gates, Memory, and Readout of a Trapped-Ion Quantum Bit, Physical Review Letters 113, 220501 (2014).

\bibitem{Ballance2016}
C. J. Ballance, T. P. Harty, N. M. Linke, M. A. Sepiol, and D. M. Lucas, High-Fidelity Quantum Logic Gates Using Trapped-Ion Hyperfine Qubits. Physical Review Letters 117, 060504 (2016).

\bibitem{Bermudez2017}
A. Bermudez, X. Xu, R. Nigmatullin, J. O’Gorman, V. Negnevitsky, P. Schindler, T. Monz, U. G. Poschinger, C. Hempel, J. Home, F. Schmidt-Kaler, M. Biercuk, R. Blatt, S. Benjamin, and M. Müller, Assessing the Progress of Trapped-Ion Processors Towards Fault-Tolerant Quantum Computation, Physical Review X 7, 041061 (2017)

\bibitem{Linke2017}
N. M. Linke, D. Maslov, M. Roetteler, S. Debnath, C. Figgatt, K. A. Landsman, K. Wright, and C. Monroe, Experimental comparison of two quantum computing architectures, PNAS 114, 3305 (2017).

\bibitem{Bruzewicz2019}
C. D. Bruzewicz, J. Chiaverini, R. McConnell, and J. M. Sage, Trapped-Ion Quantum Computing: Progress and Challenges, arXiv:1904.04178 (2019).

\bibitem{Wright2019}
K. Wright, K. M. Beck, S. Debnath, J. M. Amini, Y. Nam, N. Grzesiak, J. S. Chen, N. C. Pisenti, M. Chmielewski, C. Collins, K. M. Hudek, J. Mizrahi, J. D. Wong-Campos, S. Allen, J. Apisdorf, P. Solomon, M. Williams, A. M. Ducore, A. Blinov, S. M. Kreikemeier, V. Chaplin, M. Keesan, C. Monroe, J. Kim, Benchmarking an 11-qubit quantum computer, arXiv:08181 (2019)

\bibitem{Roos2008}
C. F. Roos, Ion trap quantum gates with amplitude-modulated laser beams, New Journal of Physics 10, 1 (2008).

\bibitem{Haddadfarshi2016}
F. Haddadfarshi and F. Mintert, High fidelity quantum gates of trapped ions in the presence of motional heating, New Journal of Physics, 18 (2016).

\bibitem{Palmero2017}
M. Palmero, S. Martinez-Garaot, D. Leibfried, D. J. Wineland, and J. G. Muga, Fast phase gates with trapped ions, Physical Review A 95, 022328 (2017).

\bibitem{Manovitz2017}
T. Manovitz, A. Rotem, R. Shaniv, I. Cohen, Y. Shapira, N. Akerman, A. Retzker, and R. Ozeri, Fast Dynamical Decoupling of the Mølmer-Sørensen Entangling Gate, Physical Review Letters 119, 220505 (2017).

\bibitem{Wong2017}
J. D. Wong-Campos, S. A. Moses, K. G. Johnson, and C. Monroe, Demonstration of Two-Atom Entanglement with Ultrafast Optical Pulses, Physical Review Letters 119, 230501 (2017)

\bibitem{Schafer2018}
V. M. Schäfer, C. J. Ballance, K. Thirumalai, L. J. Stephenson, T. G. Ballance, A. M. Steane, and D. M. Lucas, Fast quantum logic gates with trapped-ion qubits, Nature 555, 75 (2018).

\bibitem{Leung2018}
P. H. Leung, K. A. Landsman, C. Figgatt, N. M. Linke, C. Monroe, and K. R. Brown, Robust 2-Qubit Gates in a Linear Ion Crystal Using a Frequency-Modulated Driving Force, Physical Review Letters 120, 020501 (2018).

\bibitem{Webb2018}
A. E. Webb, S. C. Webster, S. Collingbourne, D. Bretaud, A. M. Lawrence, S. Weidt, F. Minter, and W. K. Hensinger, Resilient Entangling Gates for Trapped Ions, Physical Review Letters 121, 180501 (2018).

\bibitem{Shapira2018}
Y. Shapira, R. Shaniv, T. Manovitz, N. Akerman, and R. Ozeri, Robust Entanglement Gates for Trapped-Ion Qubits, Physical Review Letters 121, 180502 (2018).

\bibitem{Figgatt2018}
C. Figgatt, A. Ostrander, N. M. Linke, K. A. Landsman, D. Zhu, D. Maslov, C. Monroe, Parallel Entangling Operations on a Universal Ion Trap Quantum Computer, arXiv:1810.11948 (2018).

\bibitem{Milne2018}
A. R. Milne, C. L. Edmunds, C. Hempel, F. Roy, S. Mavadia, M. J. Biercuk, Phase-modulated entangling gates robust to static and time-varying errors, arXiv:1808.10462 (2018).

\bibitem{Leung2018b}
P. H. Leung and K. R. Brown, Entangling an arbitrary pair of qubits in a long ion crystal, Physical Review A 98, 032318 (2018).

\bibitem{Grzesiak2019}
N. Grzesiak, R. Blumel, K. Beck, K. Wright, V. Chaplin, J. M. Amini, N. C. Pisenti, S. Debnath, J. Chen and Y. Nam, Efficient Arbitrary Simultaneously Entangling Gates on a trapped-ion quantum computer, arXiv:1905.09294

\bibitem{Sutherland2019}
R. T. Sutherland, R. Srinivas, S. C. Burd, D. Leibfried, A. C. Wilson, D. J. Wineland, D. T. C. Allcock, D. H. Slichter and S. B. Libby, Versatile laser-free trapped-ion entangling gates, New Journal of Physics 21, 033033 (2019).

\bibitem{Blumel2019}
R. Blumel, N. Grzesiak and Y. Nam, Power-optimal, stabilized entangling gate between trapped-ion qubits, arXiv:1905.09292 (2019).

\bibitem{Lu2019}
Y. Lu, S. Zhang, K. Zhang, W. Chen, Y. Shen, J. Zhang, J. Zhang, and K. Kim, Scalable global entangling gates on arbitrary ion qubits, arXiv:1901.03508 (2019).

\bibitem{Sutherland2019b}
R. T. Sutherland, R. Srinivas, S. C. Burd, H. M. Knaack, A. C. Wilson, D. J. Wineland, D. Leibfried, D. T. C. Allcock, D. H. Slichter,and S. B. Libby, Laser-free trapped-ion entangling gates with simultaneous insensitivity to qubit and motional decoherence, arXiv:1910.14178 (2019).

\bibitem{Monroe2013}
C. Monroe, J. Kim, Scaling the Ion Trap Quantum Processor, Science 339, 1164 (2013).

\bibitem{Martinez2016}
E. A. Martinez, T. Monz, D. Nigg, P. Schindler, and R. Blatt, Compiling quantum algorithms for architectures with multi-qubit gates, New Journal of Physics 18, 063029 (2016)

\bibitem{Maslov2018}
D. Maslov and Y. Nam, Use of global interactions in efficient quantum circuit constructions, New Journal of Physics 20, 033018 (2018)

\bibitem{Poras2004}
D. Porras and J. I. Cirac, Effective Quantum Spin Systems with Trapped Ions, Physical Review Letters 92, 207901 (2004).

\bibitem{Islam2013}
R. Islam1, C. Senko, W. C. Campbell, S. Korenblit, J. Smith, A. Lee, E. E. Edwards, C. C. J. Wang, J. K. Freericks and C. Monroe, Emergene and Frustration of Magnetism with Variable-Range Interactions in a Quantum Simulator, Science 340, 583 (2013).

\bibitem{Jurcevic2017}
P. Jurcevic, H. Shen, P. Hauke, C. Maier, T. Brydges, C. Hempel, B. P. Lanyon, M. Heyl, R. Blatt, and C. F. Roos, Direct Observation of Dynamical Quantum Phase Transitions in an Interacting Many-Body System, Physical Review Letters 119, 080501 (2017).

\bibitem{Zhang2017}
J. Zhang, G. Pagano, P. W. Hess, A. Kyprianidis, P. Becker, H. Kaplan, A. V. Gorshkov, Z. X. Gong and C. Monroe, Observation of a many-body dynamical phase transition with a 53-qubit quantum simulator, Nature 551, 601 (2017).

\bibitem{Monz2011}
T. Monz, P. Schindler, J. T. Barreiro, M. Chwalla, D. Nigg, W. A. Coish, M. Harlander, W. Hänsel, M. Hennrich, and R. Blatt, 14-Qubit Entanglement: Creation and Coherence, Physical Review Letters 106, 130506 (2011)

\bibitem{Ozaeta2019}
A. Ozaeta, and P. L. McMahon, Decoherence of up to 8-qubit entangled states in a 16-qubit superconducting quantum processor, Quantum Science and Technology 4, 025015 (2019).

\bibitem{Greenberger1989}
D. M. Greenberger, M. A. Horne and A. Zeilinger, Going Beyond Bell's Theorem, arXiv:0712.0921 (1989).

\bibitem{Gottesman2019}
D. Gottesman, Maximally Sensitive Sets of States, arXiv:1907.05950.

\bibitem{Molmer11999}
K. M\o lmer and A. S\o rensen, Quantum Computation with Ions in Thermal Motion, Physical Review Letters 82, 1971 (1999).

\bibitem{Sorensen2000}
A. S\o rensen and K. M\o lmer, Entanglement and quantum computation with ions in thermal motion, Physical Review A, 62(2) 022311 (2000).

\bibitem{Su1980}
W. P. Su, J. R. Schrieffer, and A. J. Heeger, Soliton excitations in polyacetylene, Physical Review B 22, 2099 (1980).

\bibitem{Molmer21999}
K. M\o lmer and A. S\o rensen, Multiparticle Entanglement of Hot Trapped Ions, Physical Review Letters 82, 1835 (1999).

\bibitem{Magnus1954}
W. Magnus, On the exponential solution of differential equations for a linear operator, Communications on Pure Applied Mathematics VII, 649 (1954).

\bibitem{Blanes2009}
S. Blanes, F. Casas, J. A. Oteo, and J. Ros, The Magnus expansion and some of its applications, Physics Reports 470, 5 (2009).

\bibitem{Garey1979}
M. R. Garey and D. S. Johnson, Computers and Intractability: A Guide to the Theory of NP-Completeness, W. H. Freeman (1979).

\bibitem{Grenet2010}
B. Grenet, P. Koiran, and N. Portier, The Multivariate Resultant Is NP-hard in Any Characteristic, Part of the Lecture Notes in Computer Science book series, 6281 (2010).

\bibitem{Johanning2016}
M. Johanning, Isospaced linear ion strings, Applied Physics B 122, 71 (2016).

\bibitem{James1998}
D. F. V. James, Quantum dynamics of cold trapped ions with application to quantum computation, Applied Physics B 66, 181 (1998).

\end{thebibliography}
\end{document}